\documentclass[aps,pre,twocolumn,showpacs,superscriptaddress]{revtex4-1}
\usepackage{float}
\usepackage{lineno}
\usepackage{lipsum}
\usepackage{mathrsfs}
\usepackage{graphicx}
\usepackage{wrapfig}
\usepackage{bm}
\usepackage{amsmath,amssymb}
\usepackage{mwe}   
\usepackage{subfig}
\usepackage{color}
\usepackage[table]{xcolor}
\definecolor{truegreen}{RGB}{8, 148, 4}
\definecolor{irishgreen}{rgb}{0, 0.565, 0.165}
\usepackage[colorlinks = true,linkcolor=black, citecolor=irishgreen,urlcolor =irishgreen]{hyperref}
\usepackage{stackengine}
\usepackage[font=normalsize] {caption}
\usepackage[section]{placeins}
\usepackage[english]{babel} 

\makeatletter
\newcommand*{\rom}[1]{\expandafter\@slowromancap\romannumeral #1@}
\makeatother

\renewcommand{\r}{{\bm r}}

\begin{document}
\title{Clustering in quasi-two-dimensional dispersions of Brownian particles with competitive interactions: Phase diagram and structural properties}
\author{Zihan Tan}\email{zihan.tan@tu-berlin.de}
\affiliation{Biomacromolecular Systems and Processes (IBI-4), Institute of Biological Information Processing, Forschungszentrum J\"ulich, 52428 J\"ulich, Germany}
\affiliation{Computational Biomedicine (IAS-5/INM-9), Institute for Advanced Simulation, Forschungszentrum Jülich, 52428 Jülich, Germany}
\affiliation{Present address: Institut f\"ur Theoretische Physik, Technische Universit\"at Berlin, Hardenbergstr$\beta$e 36,10623 Berlin, Germany}
\author{Vania Calandrini}
\affiliation{Computational Biomedicine (IAS-5/INM-9), Institute for Advanced Simulation, Forschungszentrum Jülich, 52428 Jülich, Germany}
\author{Jan K. G. Dhont}
\affiliation{Biomacromolecular Systems and Processes (IBI-4), Institute of Biological Information Processing, Forschungszentrum J\"ulich, 52428 J\"ulich, Germany}
\affiliation{Department of Physics, Heinrich-Heine Universit\"at D\"usseldorf, D-40225 D\"usseldorf, Germany}
\author{Gerhard N\"agele}
\affiliation{Biomacromolecular Systems and Processes (IBI-4), Institute of Biological Information Processing, Forschungszentrum J\"ulich, 52428 J\"ulich, Germany}
\affiliation{Department of Physics, Heinrich-Heine Universit\"at D\"usseldorf, D-40225 D\"usseldorf, Germany}
\begin{abstract}
Competing short-range attractive (SA) and long range repulsive (LR) interactions have been invoked to describe colloid or protein solutions, as well as membrane proteins interactions mediated by lipid molecules. Using Langevin dynamics simulations, we determine the generalized phase diagram, the cluster shapes and size distributions of a generic Q2D dispersion of spherical SALR particles confined to in-plane motion. SA and LR interactions are modelled by a generalized Lenard-Jones potential and a screened Coulomb potential, respectively.  The microstructures of the various equilibrium and non-equilibrium phases turn out to be distinctly different from the ones observed in three-dimensional (3D) SALR systems. We discuss perturbation theory predictions for the metastable binodal line of a reference system of particles with SA interactions only, which in the Q2D-SALR phase diagram separates cluster from non-cluster phases. The transition from the high-temperature (low SA) dispersed fluid phase to the lower-temperature equilibrium cluster phase is characterised by a low-wavenumber peak of the static structure factor (corresponding to a thermal correlation length of about twice the particle diameter) featuring a distinctly smaller height ($\approx1.4$) than in 3D SALR systems. By further decreasing the temperature (increasing SA), the cluster morphology changes from disk-like shapes in the equilibrium cluster phase, to double-stranded anisotropic hexagonal cluster forms in the cluster-percolated gel phase. This transition is quantified by the hexagonal order parameter distribution function. The mean cluster size and coordination number of particles in the gel phase are insensitive to changes in the attraction strength.
 
\end{abstract}

\maketitle

\section{Introduction}
Over the past two decades, immense attention has been paid to Brownian particles interacting via competitive short-range attractive (SA) and long-range repulsive (LR) forces~\cite{strad:2004,yliu:2019,ruiz:2021}. The competition between attraction and repulsion of these particles makes them excellent model system for understanding the physics of a plenteous amount of soft matter and biological systems, such as charged colloids with doped depletants~\cite{strad:2004,sedg:2004,camp:2005prl,klix:2010,kohl:2016,redd:2012}, lysozyme proteins at low-salinity conditions~\cite{strad:2004,cardinaux:2007,yliu:2011,cardinaux:2011,godfrin:2015,godfrin:2018,ries:2018}, Y-shaped monoclonal antibodies~\cite{year:2014,godf:2016,Chowdhury:2023}, and the reentrant liquid condensation of Ribonucleoprotein-RNA complexes~\cite{Alshareedah:2019} to name a few. Practically, querying the respective influence of SA and LR provides means to control the microstructure of solutions and is also of importance for applications such as in the formulation of pharmaceutical drugs and protein crystallization~\cite{yliu:2019}. Owing to its values for both fundamental theory and industrial applications, the structure and phase behavior of dispersion with SALR pair potential in three-dimensions have been exhaustively investigated analytically, numerically and experimentally. For example, integral equation theory~\cite{pini:2006,foffi:2002}, discrete perturbation theory (DPT)~\cite{benavides:1999,jimenez2D:2008,godfrin:2014}, mean-field density functional theory (DFT)~\cite{archer:2007, chacko:2015,archer:2007pre,archer:2008pre,riest2015salr}, and computer simulations~\cite{godfrin:2014,das2018clustering} have been used to study the equilibrium phase diagrams of 3D SALR systems. Experimentally, SALR pair interactions have been invoked to explain the behavior of lysozyme solutions at low salts concentrations\cite{godfrin:2018,ries:2018}.\\
\indent Contrasting to bare attractive particles resulting in gas-liquid phase separation, the competitive repulsive interaction ingredient in SALR suspensions primarily stints the size of growing clusters, thus frustrating the phase separation (modulated phases)~\cite{bolli:2016,andelman:1999,andelman:2009}. Given its separated lengthscale of attraction and repulsion, rich phase behaviors, including equilibrium and non-equilibrium cluster phases, have been observed in three-dimensional (3D)-SALR systems~\cite{godfrin:2014,serna:2021}. 
This type of geometrically frustrated aggregations~\cite{grason:2016,selinger:2021} in SALR systems is termed  as ``intermediate range order”  (IRO)~\cite{yliu:2011} and is reflected in the static structure factor, $S(q)$, by a ``prepeak", prior to the main peak associated to the first neighbor cage. This prepeak $S(q_c)$ in 3D-SALR is closely bundled with the onset of cluster formation and has been intensively debated~\cite{yliu:2019,yliu:2011}.\\
\indent Particularly, a notable effort has been devoted to formulating universal criteria that classify the modulated (micro-) phase separation in 3D-SALR systems. Commonly, empirical criteria used to identify macrophase separation established in purer and simpler interactions (e.g., hard-sphere systems) are analogously compared with more complex SALR systems. Godfrin and co-workers explored a wide range of state points and obtained the generalized phase diagram. Albeit the universality is not ensured in SALR systems with relatively long-ranged repulsion~\cite{ruiz:2021}, an analogously extended law of corresponding-states (ELCS) has been established for the 3D-SALR system. The NF-ELCS, initially formulated by Noro \& Frenkel (NF)~\cite{norefrenkel:2000}, states that short-ranged (less than $25\%$ of the particle diameter) purely attractive systems share common thermodynamic behavior when compared at the same second virial coefficient. They considered the binodal curve of the purely SA fluids as an approximate indicative separating fluid and cluster phases in the SALR system~\cite{godfrin:2014}. Note that the emergence of $S(q_c)$ is always considered a prerequisite, rather than a stringent condition for the onset of a cluster phase. Godfrin et al. have suggested that in 3D-SALR systems, the \textit{critical} prepeak value $S(q_c)\approx 2.7$ indicates the fluid-cluster transition~\cite{godfrin:2014}. This empirical rule is the generalization of the Hansen-Verlet freezing norm stating that the fluid-solid transition takes place when $S(q_c)\approx2.85$~\cite{hasen:1969,hasen:1970}. Later on, Bollinger and Truskett suggested the width of $S(q_c)$, associated with the thermal correlation length $\xi_T$ through a Lorentzian fit, as an indicator of a cluster phase~\cite{jadri:2015,bolli:2016}. Combing the two criteria for 3D-SALR potentials, they advocated that $S(q_c)\approx2.7$ and $2.0\lesssim\xi_T/\sigma\lesssim3.0$ hallmark the onset of equilibrium clusters phase, with $\sigma$ the diameter of the Brownian particle. Besides, the empirical correlation between the mean cluster size and the average coordination number has also been discussed since they are associated with the enthalpic contribution and entropic reduction during cluster formation~\cite{godfrin:2014}.\\
\indent Moving to quasi-two-dimensional(Q2D)-SALR systems, it is natural to ask if cluster properties and phase behavior are similar to their 3D counterpart as well as if analogous empirical criteria apply. In a more general sense, compared with the 3D situation, distinctively different physics could appear in 2D. A celebrated example is that the 2D melting of hard spheres (disks) exhibits two continuous transitions through an intermediate hexatic phase captured by Kosterlitz–Thouless–Halperin–Nelson–Young (KTHNY) theory~\cite{halpe:1978,Nelson1979,strandburg:1988}.\\
\indent Exploring the phase transitions of (Q2D)-SALR systems is particularly relevant for colloidal (or protein) particles confined at (or close to) planar interfaces. In the biological world, diffusion and self-organization of phospholipids and membrane-proteins into finite-sized domains, such as ``lipid rafts"~\cite{simons:1997,levental:2020,sezgin:2017} and protein clusters~\cite{ramamurthi:2009,merklinger:2017,Dufourc:2008}, play a significant role in signal transduction, membrane sorting, protein processing, and virus trafficking~\cite{schuette:2017,lamerton:2021,dwek:2003}. It has been evidenced that the interplay of SA (e.g., due to lipid-mediated depletion, wetting, and hydrophobic mismatch, or direct chemical interactions between amino-acid side chains) and LR (induced, e.g., by mechanical deformation or fluctuations of the membrane) is crucial for the formation of membrane-protein clusters~\cite{Destainville2018,wasnik:2015,sieber:2007,gurry:2009,destainville:08,destainville:2008}. To this end, systematic investigation of the phase behaviors, structural and cluster properties for Q2D-SALR particles becomes indispensable. Surprisingly, despite its fundamental importance, only a few efforts have been devoted to exploring the structure and dynamics of (Q)2D-SALR systems. Some theoretical and simulation works on (Q)2D-SALR systems show that particles can self assemble into various exotic microphases either in bulk~\cite{char:2007,schwa:2010,schw:2016,bores:2015} or when trapped (pinned) by external forces~\cite{liu:2008,chen:2011,campo:2013} or confinement~\cite{chacko:2015}. \\
\indent This work uses Langevin dynamics (LD) simulations to explore the phase behavior, structural and cluster properties of a monolayer of SALR Brownian particles restricted in-plane. We systematically investigate a broad range of attraction strength and density values and discuss the empirical critical criteria distinguishing disperse fluid and equilibrium clusters phases. The LD simulation results enable us not only to look into the static properties of our Q2D-SALR systems, but also to gain insights on the dynamics of cluster formation, which will be discussed extensively in a subsequent paper\cite{tan05}.\\
\indent The present paper is prepared as follows. Sec.~\ref{sec:method} outlines the LD scheme for our Q2D-SALR Brownian particles and recaps typical functions characterizing the in-plane cluster properties. Sec.~\ref{sec:results} reports the main results of this paper. Notably, the (generalized) phase diagram is obtained by analyzing the shape of the cluster size distribution function. The (generalized) phase diagram is then compared with the metastable binodal curves of purely SA square-well (SW) fluids obtained by second-order perturbation theory (PT). Furthermore, the intermediate range order in $S(q)$ is discussed, and associated empirical rules reflected in the structure factor are proposed as in the 3D case. The morphology of the clusters with hexagonal features and local orders of clusters are discussed in detail. Finally, Sec.~\ref{sec:conclusion} comprises a summary and conclusions of our observations.
\section{Methods}
\label{sec:method}
	\subsection{Algorithm}
	We consider $N$ spheres with identical mass $M$ and radius $R$ confined into a \textit{monolayer} undergoing \textit{under-damped} Brownian motion characterized by the Langevin Equation~\cite{dhont1996introduction,allen}
	\begin{equation}
		M_i\ddot{\r}_i(t)+\gamma\dot{\r}_i(t)=-\nabla_i V(t)+\bm{\Gamma}_i(t).
		\label{lgeq}
	\end{equation}
	Here, $\r_i(t)$, $\dot{\r}_i(t)$ and $\ddot{\r}_i(t)$ denote the central position, velocity, and acceleration of the $i^{\text{th}}$ particle with $i\in\{1,..,N\}$, and $\gamma$ is the translational friction coefficient, $\bm{\Gamma}_i(t)$ the Gaussian white-noise random force with moments
	\begin{equation}
		\langle\bm{\Gamma}_i(t)\rangle_\tau=0,
		\label{eq:meanNoise}
	\end{equation}
	and
	\begin{equation}
		\langle\bm{\Gamma}_i(t)\bm{\Gamma}_j(t')\rangle_\tau=2\gamma k_BT\hat{\bm{I}}\delta(t-t').
		\label{eq:varNoise}
	\end{equation}
	\indent In Eq.~\eqref{eq:varNoise}, $\langle\cdots\rangle_\tau$ encodes the average over time, $k_B$ denotes the Boltzmann constant, $T$ the equilibrium temperature, $\hat{\bm{I}}$ the identity matrix. Note that the isolated particle self diffusion coefficient $D_0$ is related to $\gamma$ via the fluctuation dissipation theorem, namely $D_0=k_BT/\gamma$. Further, the noise term acts as a thermostat that keeps the temperature of the system at $T$. The force acting on each particle $i$ resulting from the SALR interactions with all the other particles is $-\nabla_i V(t)=\sum_{i\neq j}^{N}\frac{\partial U(r_{ij})}{\partial r_{ij}}$. Here $V(t)$ is the total potential of the system, which is the sum of all pair-wise potential $U(r_{ij})$ with $r_{ij}$ the center-to-center distance between particle $i$ and $j$. In this work, the SALR potential is given by~\cite{char:2007,mani:2014,das2018clustering}
	\begin{equation}	\label{eq:ljy}
		U\left(r_{ij}\right) = \left\{\begin{tabular}{l}
			$\infty;$  $r \le \sigma$ \\
			$4\epsilon\left[ \left(\frac{\sigma}{r_{ij}}\right)^{100}-\left(\frac{\sigma}{r_{ij}}\right)^{50} \right]$\\$+\ell_B Z_{\text{eff}}^2\frac{e ^{-r_{ij}/\lambda}}{r_{ij}};$ 
			$\sigma < r \le r_\mathrm{c}$\\
			$0;$ $r> r_\mathrm{c}$
		\end{tabular}\right.
	\end{equation}
	The first term in the r.h.s. of Eq.~\eqref{eq:ljy} is a generalized 100-50 Lennard-Jones potential with a hard-sphere-like steep repulsion part and a short-range attraction part (only a few percent of the particle diameter), where $\sigma=2R$ is the particle diameter and $\epsilon$  the attraction strength (depth). The second term is a long-range, repulsive screened Coulomb (Yukawa) potential, where $\ell_B$ is the Bjerrum length of the solvent, $Z_{\text{eff}}$ is effective particle valency and $\lambda$ the Debye-H\"{u}ckel screening length due to counter- and electrolyte ions dissolved in the solvent. In simulations, the potential is truncated with a shift at $5\sigma$~\cite{mani:2014}. We emphasize that albeit the centers of the diffusing particles are confined on a plane, the screened Coulomb potential in Eq.~\eqref{eq:ljy} is formulated in 3D to mimic the Q2D conditions.
	
	Furthermore, in order to calculate the liquid-gas coexisting line (binodal curve), we simulate a reference system where only the SA component is active. The cutoff of the pure SA potential is set to the second zero-crossing point $x_0$ of the potential in Eq.~\eqref{eq:ljy}~\cite{godfrin:2014}. Thus, the attraction range $x_0 (\epsilon)$ varies as the attraction strength $\epsilon$ changes. One can employ the 2D $2^{nd}$ order (discrete) perturbation theory to calculate the binodal~\cite{godfrin:2014, refId:2008,chapela:2010,torres-arenas:2010,valadez-perez:2012}. However, to avoid the cumbersome numerical calculation introduced by the complex form of Eq.~\eqref{eq:ljy}, we approximate the SA system by an attractive square well (SW) fluid with a similar attraction range, expressed as
	\begin{equation}	\label{eq:sqw}
		U_{sw}\left(r_{ij}\right) = \left\{\begin{tabular}{l}
			$\infty;   r_{ij}\le\sigma$ \\
			$-\epsilon; \sigma < r \le \lambda\sigma$\\
			$0; r> \lambda\sigma$
		\end{tabular}\right.
	\end{equation}
	Here, $\lambda>1$ denotes the attraction range. In our simulations, the values of attraction strength $\epsilon$ in the range of $\left[2, 20\right]k_BT$ are covered. The mean cut-off distance of the according reference SW systems is then given by  $\overline{x_0(\epsilon)}/\sigma=\lambda_m\approx1.06$.  
	
	The binodal lines are then calculated using the 2nd order perturbation theory proposed by Trejos~\cite{trejos:2018} in the framework of the statistical associating fluid theory for potentials of variable range (SAFT-VR)~\cite{martinez:2007,trejos:2014,trejos_:2018}. The perturbation theory, in a nutshell, computes the thermodynamic quantities by treating the potential of the interacting system as the perturbation of the hard-disk (hard-sphere in 3D) potential. In our calculations, the Helmholtz free energy is expanded to the 2nd order term.
	
	Here and in what follows, the energy is expressed in unit of $k_BT$, length scales are considered in unit of $\sigma$, time scales in unit of $\sigma\sqrt{M/k_BT}$. We fix $ \ell_B Z^2_{\text{eff}}=3.588\sigma$ and $ \lambda=1.794\sigma$ in all the
	calculations reported in this paper. The value of the friction coefficient of each particle is chosen as $\gamma=80.1 \sqrt{Mk_BT}/\sigma$ leading to the one-particle momentum relaxation time $\tau_\gamma=M/\gamma=0.0125\sigma\sqrt{M/k_BT}$. Moreover, hereinafter, an effective temperature $T^*=k_BT/\epsilon$ is introduced to characterize the influence of the attraction strength. With this parameterization, the attraction range of the interaction potential in SALR particles is less than $20\%$ of the particle diameter whist the repulsive range ($\lambda$) is comparable than to the one used in previous simulation works on 3D-SALR systems~\cite{godfrin:2014, yliu:2019,sciortino:2004}. 
	\subsection{Cluster classification}
	To identify its phase behavior, we analyze the cluster size distribution function (CSD), $N(s)$, calculated from numerical simulations. $N(s)$ is defined as
	\begin{equation}
		N(s)=\left<\frac{s\cdot n_c(s)}{N}\right>,
		\label{eq:csd}
	\end{equation}
	denoting the mean fraction of particles involved in clusters of size $s$ (i.e. containing $s$ particles). Here, $\langle\cdots\rangle$ is the average over representative particle configurations, $n_c(s)$ is the number of clusters of size $s$ within a given configuration. The sum of $N(s)$ obeys uniformity condition $\sum^N_{s=1}N(s)=1$. Whether two interacting particles belong to the same cluster is determined by their mutual distance $r_{ij}$, which has to be not greater than the threshold value , $x^*$, where the potential in Eq.~\eqref{eq:ljy} is at the local maximum (potential barrier). This is reasonable because when $r_{ij}\leq x^*$ the particle tends to slide down towards the attraction well.
	\subsection{Hexagonal order parameter}
	Particles with SALR interactions in 2D tend to form hexagonal clusters~\cite{schw:2016}. The hexagonal order can be identified by the hexagonal orientational order parameter $\left|q_6\right|^2$~\cite{halpe:1978,Nelson1979,bialk:2012,zottl2014,theers2018clustering,ruiz:2019}. For particle $i$, it is defined as
	\begin{equation}
		q^i_6=\frac{1}{6}\sum_{j\in N_i^{(6)}}\exp\{i6\alpha_{ij}\},
		\label{eq:hex}
	\end{equation}
	where the sum runs over the six (intra- or inter-cluster) nearest neighbors of particle $i$ and $\alpha_{ij}$ is the angle between $r_{ij}$ and the $x-$axis. For a perfect hexatic lattice $\left|q^i_6\right|^2=1$. For brevity, the index of the particle will be omitted.
	\subsection{Simulation setup}
	We explore the structure and phase behavior of different Q2D-SALR systems by examining over $8$  values of effective temperature $T^*=k_BT/\epsilon$ under $4$ different conditions of particle density (per area concentration) $\phi_{2D}$. We fix the number of particles while varying the simulation box length to reach distinct density values. For each parameter combination $(\phi_{2D}, T^*)$, all quantities are measured after performing LD simulation over a sufficiently long time to let the system reach the stationary state. If not specified, the results shown below are for $N=1024$ particles. For some specific parameter conditions, larger systems have been considered to check the consistency of the results.
\section{Results}
\label{sec:results}
\begin{figure}[h]
	\subfloat{		
		\begin{picture}(100,180)
			\put(-72,5){\includegraphics[width=0.46\textwidth,trim={0 0 0 0},clip]{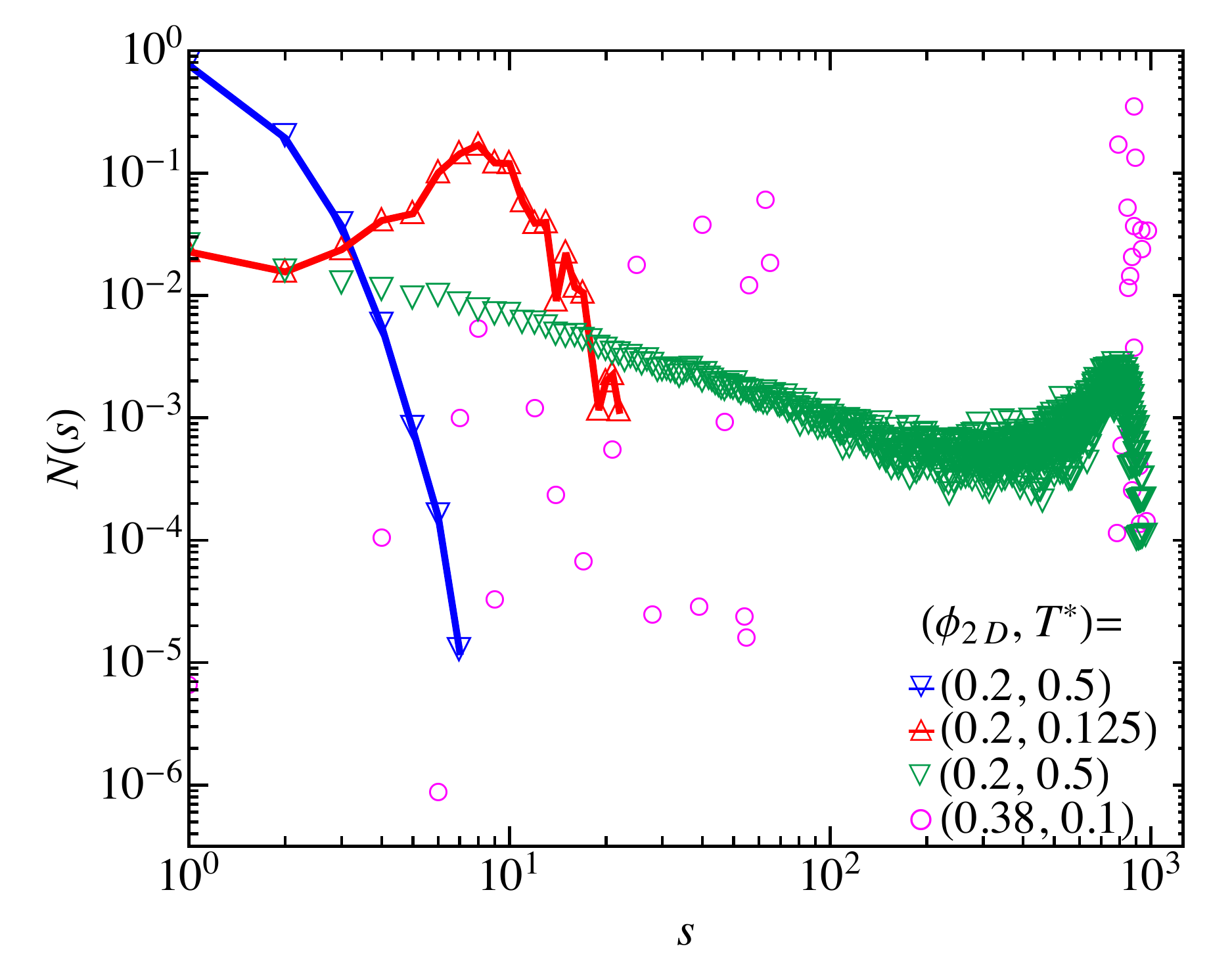}}
			\put(130,165){(a)}
		\end{picture}
		\label{fig:csd}
	}\\
	\vspace{-5mm}
	\subfloat{		
		\begin{picture}(100,180)
			\put(-80,-10){\includegraphics[width=0.48\textwidth,trim={0 0 0 0},clip]{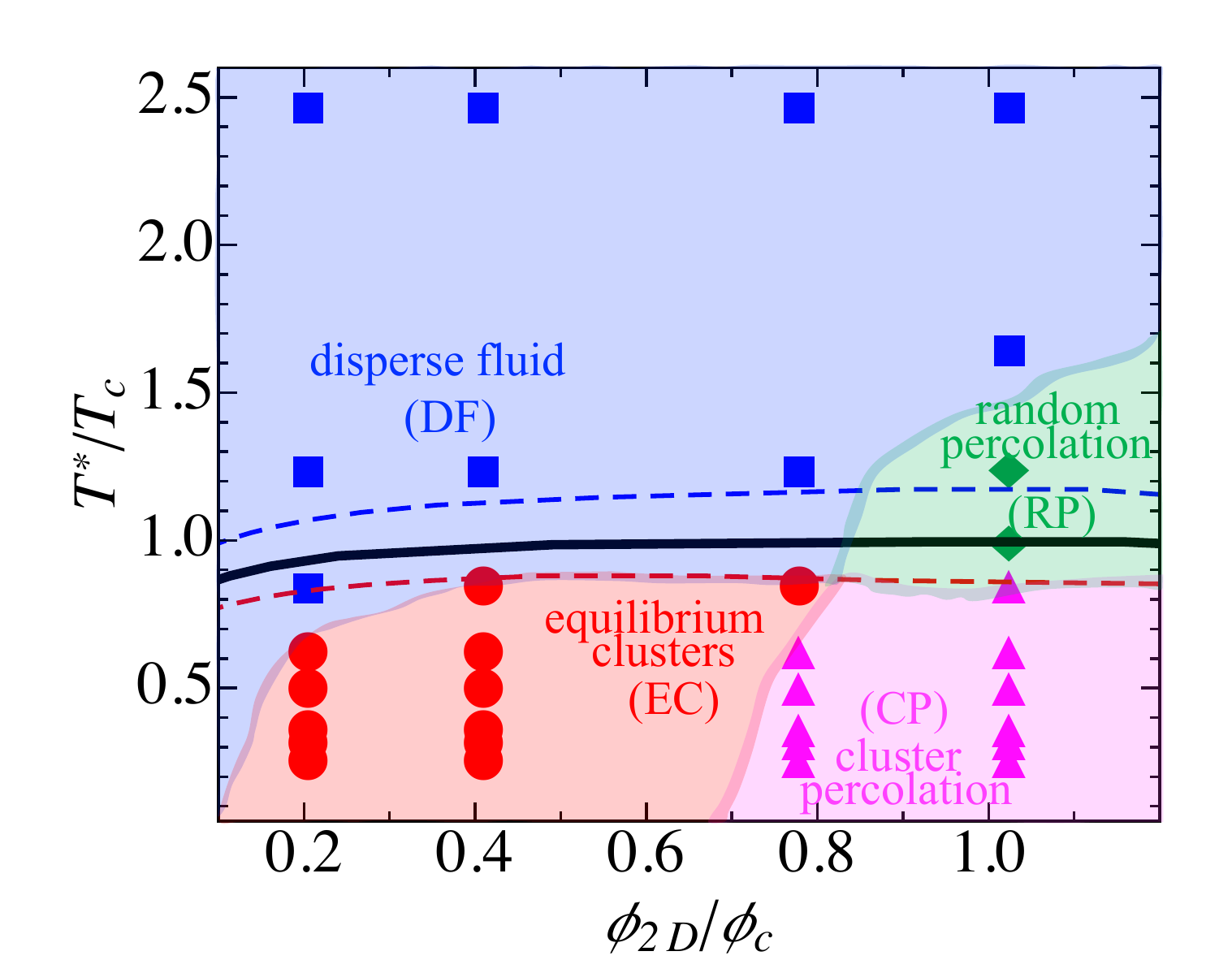}}
			\put(130,155){(b)}
		\end{picture}
		\label{fig:std}
	}
	\vspace{3mm}
	\caption{\label{fig:phases}(a)~Cluster size distribution function, $N(s)$, of Q2D-SALR systems for $(\phi_{2D}, T^*)$ values 
		corresponding to \textit{dispersed fluid} (blue down triangles), \textit{equilibrium clusters} (red up triangles), \textit{random percolation} (green diamonds), and \textit{cluster percolation} (magenta circles) phases. (b)~Q2D-SALR phase diagram. All $34$ state points, falling into four categories, are marked in different symbols and colors. The lines correspond to the binodal coexisting curves of square-well (SW) fluids with attraction range parameter $\lambda=1.075$ (dashed blue), $1.06$ (continuous black) and $1.05$ (dashed red), computed by the 2nd order perturbation theory proposed by Trejos et al. \cite{trejos:2018}. All of the state points $(\phi_{2D}, T^*)$ are normalized by the values at the critical point$(\phi_c, T_c)$ of the SW fluids for $\lambda=1.06$.}
\end{figure}
\begin{figure}
	\raggedright
\includegraphics[height=0.45\textwidth]{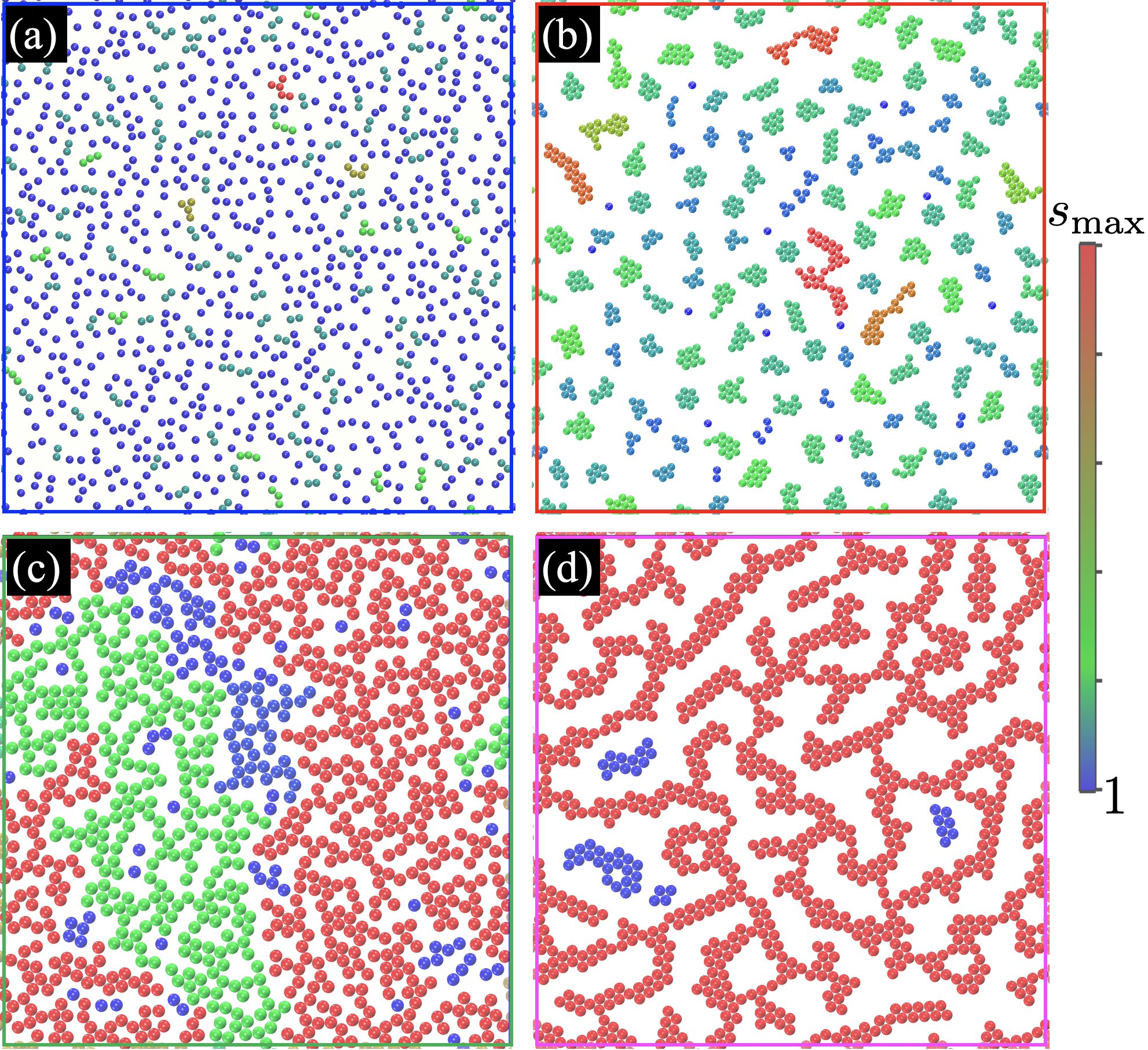}
\caption{\label{fig:snapshots}Typical snapshots of Q2D-SALR systems in different phases (marked by colored frames): (a)~dispersed fluid for $( \phi_{2D},T^*)=(0.2, 0.5)$, (b)~equilibrium clusters for $( \phi_{2D},T^*)=(0.2, 0.125)$, (c)~random percolation for $( \phi_{2D},T^*)=(0.5, 0.25)$ and (d)~cluster percolation for $( \phi_{2D},T^*)=(0.38, 0.1)$. The particle color code represents clusters of the same size $s$, ranging from $1$ (monomer) in blue to the maximum size $s_{\text{max}}$ in red. Note that the $4$ density conditions correspond to simulation boxes of different size, since $N=1024$ is fixed. The representative  snapshots shown in the panels are rescaled for graphical reasons.}
\end{figure}
	\subsection{Generalized phase diagram}
As intensively discussed for 3D-SALR systems, the shape of the $N(s)$ curve is considered as an indicator of the system phase. In the same vein, we characterize the phase behavior by computing the CSD $N(s)$ (cf. Eq.~\eqref{eq:csd}).  Fig~\ref{fig:phases}\subref{fig:csd} presents the $N(s)$ for different $(\phi_{2D}, T^*)$ values, which identify four characteristic phases, viz.: dispersed fluid (blue down triangles, $(\phi_{2D}, T^*)=(0.2, 0.5)$) exhibiting a monotonous decay of $N(s)$, equilibrium clusters (red up triangles, $(0.2, 0.125)$) with a local maximum (finite cluster peak) in $N(s)$, random percolation (green triangles, $(0.5, 0.25)$) where a cluster peak arises at $s\approx N$ after a monotonical decay, and cluster percolation (magenta circles, $(0.38, 0.1)$) showcasing a percolated system-spanning cluster peak and additional oscillating intermediate peaks (here, $s\approx 50$). This criterion is used to identify the phase of all the simulated state points and build the $(\phi_{2D}, T^*)$ phase diagram shown in Fig.~\ref{fig:phases}\subref{fig:std}. The phase diagram resembles that in 3D-SALR case. At relatively high effective temperature $T^*$ and low $\phi_{2D}$, the Q2D-SALR system maintains dispersed fluid phase. Equilibrium clusters and cluster percolated phases are observed at lower $T^*<0.25$ and higher $\phi_{2D}$. Systems with intermediate attraction strength and high enough density result in random percolation phase.

The correspondent liquid-gas coexisting curves (solid black) of attractive SW fluids (cf. Eq.~\eqref{eq:sqw}), with the attraction range parameter values in the vicinity of  $\lambda_m$ (from bottom to top, $\lambda=1.05$, $1.06$, and $1.075$, respectively) are displayed in Fig.~\ref{fig:phases}\subref{fig:std}. The three curves overall divide the fluid and clustered phases (micro-phase separation) in the phase diagram. An increase of $\lambda$ in the SW fluids leads to the coexisting curve and the critical temperature $T_c$ shifting towards higher temperature. More interestingly, akin to the observation in 3D, the line corresponding to $\lambda_m$ encloses the clustered phase in the 2D-SALR system. All the data points are represented by their normalized values with respect to the critical point of the $\lambda_m=1.06$ curve. Explicitly, $(\phi_c, T_c)=(0.488, 0.203)$.

Godfrin et al. also referred to this micro-phase separation in 3D-SALR system as an extension of the well-known Noro \& Frenkel ELCS. In contrast, there are indications that the ELCS is violated in two dimensions. Our calculations show that the reduced second virial coefficient at the critical point is highly sensitive to the attraction range parameter $\lambda$. Moreover, to the best of our knowledge, the 2D-ELCS has not been explicitly discussed. On the other hand, the 2nd order perturbation theory is less accurate in two dimensions due to more pronounced critical fluctuations. For $\lambda\le1.5$, the binodals of SW fluids obtained by the SAFT-VR approach are also shown to underestimate the flatness of the curve when compared with computer simulation results~\cite{trejos:2018}. Therefore, we could not draw such an analogous conclusion on ELCS in 2D-SALR system.\\
\indent Fig.~\ref{fig:snapshots} illustrates the (stationary) configurations of Q2D-SALR particles for four $(\phi_{2D}, T^*)$ values as in Fig.~\ref{fig:phases}\subref{fig:csd}, in which particles belonging to clusters of the same size $s$ are shown in the same color. Here, some comments are in order. First, for systems at dispersed fluid phase (cf. Fig.~\ref{fig:snapshots}(a)) dominated by monomers, small clusters with various size and loosely packed are occasionally recognized. Second, in equilibrium clusters phase (Fig.~\ref{fig:snapshots}(b))), the clusters show a preferred size (around $10$ for the specific conditions, Fig~\ref{fig:phases}\subref{fig:csd}), due to the increasing attraction (decreasing $T^*$). Finally, the random percolated phase (Fig.~\ref{fig:snapshots}(c)),  featuring spanning clusters, can be viewed as the ``crowded" dispersed fluid phase, which still lacks orientational order. In contrast, the particles at cluster percolated phase (Fig.~\ref{fig:snapshots}(d)) form compact and locally elongated structures. 

\subsection{Intermediate range order peak}
\begin{figure}[h]
	\centering
	\subfloat{		
		\begin{picture}(100,180)
			\put(-72,5){\includegraphics[width=0.46\textwidth,trim={0 0 0 0},clip]{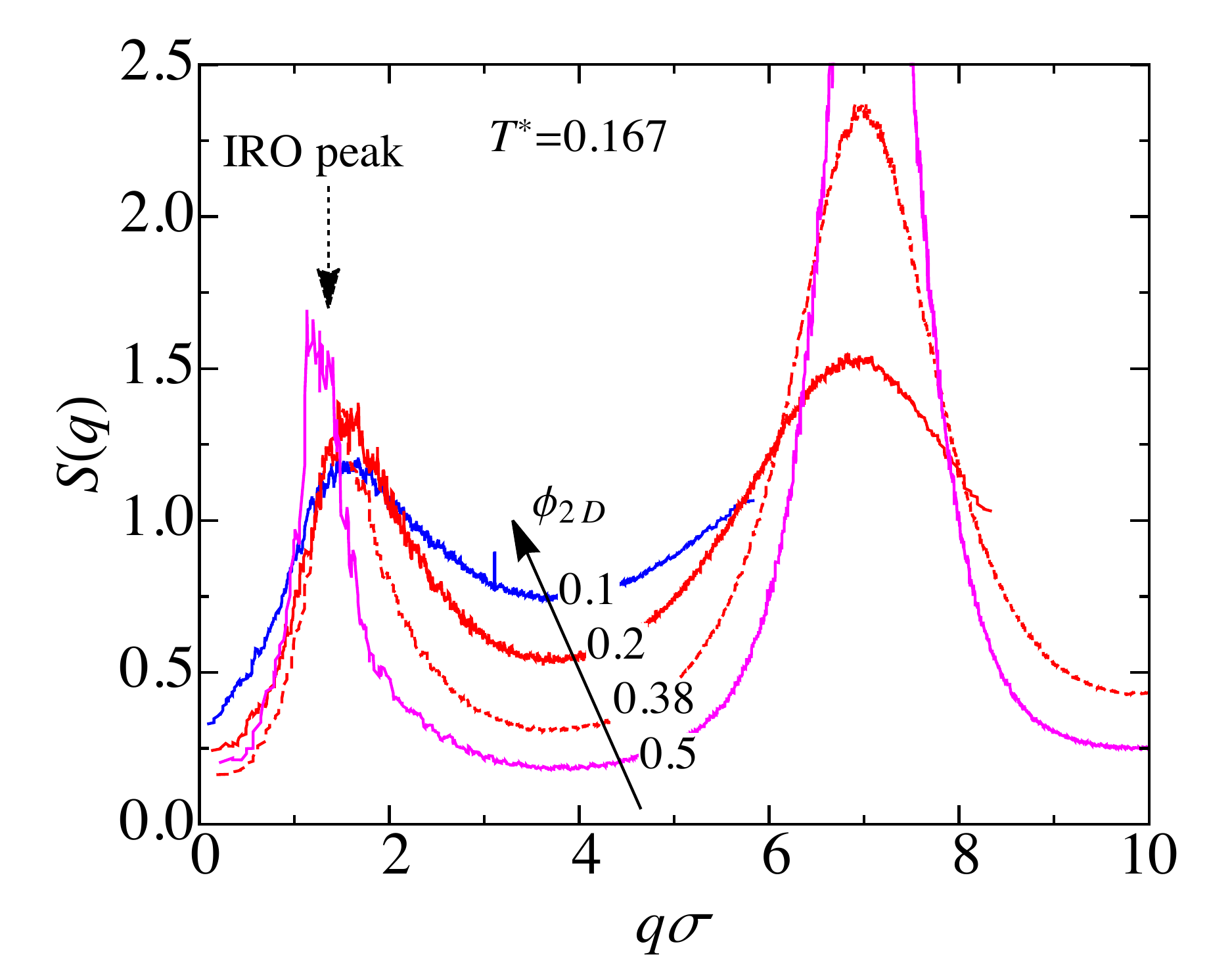}}
			\put(130,165){(a)}
		\end{picture}
		\label{fig:sq}
	}\\
	\vspace{-5mm}
	\subfloat{		
		\begin{picture}(100,180)
			\put(-80,-10){\includegraphics[width=0.5\textwidth,trim={0 0 0 0},clip]{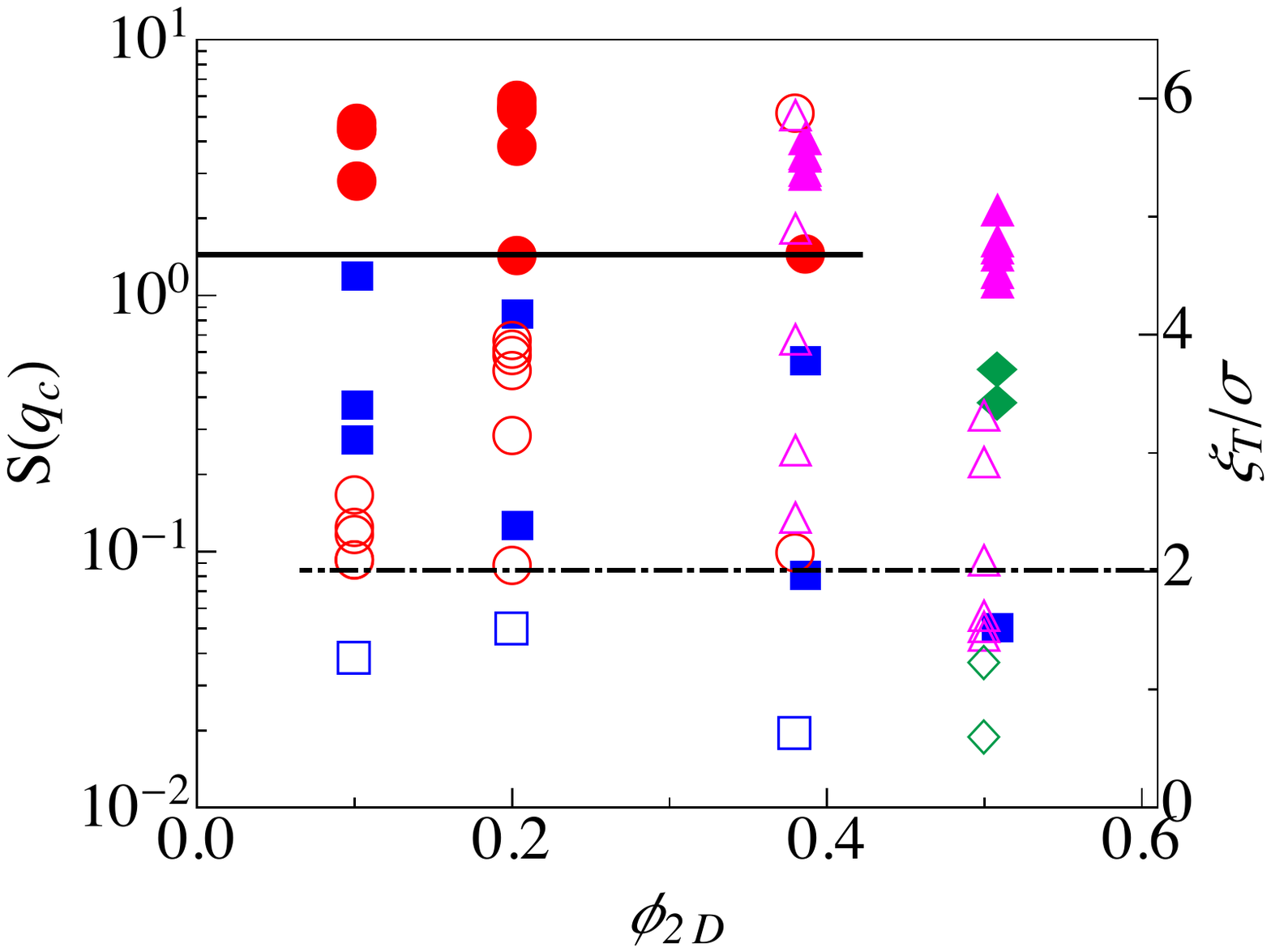}}
			\put(130,155){(b)}
		\end{picture}
		\label{fig:sq_c}
	}
	\vspace{1mm}
	\caption{(a)~Static structure factor $S(q)$ at $T^*=0.167$ ($\epsilon=6k_BT$) for area fractions as indicated. The dashed arrow marks the approximate location of the IRO peak. (b)~The static structure factor at the prepeaks $S(q_c)$ (left $y-$axis, filled symbols) and the thermal correlation length $\xi_T/\sigma$ (right $y-$axis, open symbols) in Q2D-SALR systems. The solid (dashed) line marks the critical value $S(q_c)\approx1.4$ ($\xi_T/\sigma\approx2$), separating the dispersed fluid and equilibrium clusters phases. Color code and symbols indicate different phases as in Fig.~\ref{fig:phases}\protect\subref{fig:std}.}
	\label{fig:Sqall}
\end{figure}
As in 3D-SALR systems, we aim to pin down the onset of clustering and cluster morphologies, by detecting the height of $S(q_c)$~\cite{godfrin:2014}, assisted by the thermal correlation length $\xi/\sigma$ (proportional to the width of the $S(q)$ around the prepeak)~\cite{jadri:2015,bolli:2016}. The $\xi/\sigma$ norm is analogous to quantifying the structural correlations and the macrophase liquid-gas separation, explicitly given by the second-order inverse expansion of $S(q)$ around the peak at $q_c=0$. In clustered SALR systems where frustrated interactions dominate, the inverse expansion around the prepeak is written as~\cite{bolli:2016}
\begin{equation}
	S(q\sigma)\approx\frac{S(q_c\sigma)}{1+(\xi_T/\sigma)^2(q-q_c)^2\sigma^2}.
	\label{eq:sqfit}
\end{equation}
\indent Likewise, the $S(q)$ in Q2D-SALR system demonstrates a prepeak at $S(q_c)$, before the first-neigbhbor main peak at $q_m\approx2\pi/\sigma$. Fig.~\ref{fig:Sqall}\subref{fig:sq_c} presents the $S(q)$ function for diferent area fraction at fixed temperature $T^*=0.167$. Colors of the curves code the system at corresponding phases. As increasing of $\phi_{2D}$, $S(q_c)$ increases. Interestingly, the positon of the prepeaks is roughly independent on $\phi_{2D}$. In simple random phase approximation (RPA), the direct correlation function $c(r)$ of the LJY - SALR system is expressed {\em approximately} in terms of the hard-disk direct correlation function, $c^\text{hd}(r)$.  Thus, according to the RPA in combination with assuming that $S^\text{hd}(y \approx y_c) =S^\text{hd}(y=0)$, it is predicted that $y_c$ is independent of $\phi_{2D}$ but determined only by the competing attraction / repulsion  parameters of the SALR potential. In actuallity, $y_c$ is expected to be mildly $\phi_{2D}$-dependent for fixed pair potential. Moreover, in standard RPA the perturbation potential should be weak enough that the physical condition $S(y)>0$ is fullfilled. We analogously adopt both criteria based on the IRO prepeak height and the width of $S(q_c)$, respectively, into Q2D-SALR systems. Fig.~\ref{fig:Sqall}\subref{fig:sq_c} summarizes our classification of the height of the prepeak value $S(q_c)$ (filled symbols) and the corresponding thermal correlation length $\xi_T/\sigma$ (open symbols) by the best fit based on Eq.~\eqref{eq:sqfit}. Interestingly, the observed critical values of $S(q_c)$ and $\xi_T/\sigma$ are in qualitative agreement with those in 3D. Remarkably, the Hansen-Verlet-like criterion yields a critical value $S(q_c)\approx1.4$ in (Q)2D-SALR systems instead of $2.7$ as in 3D. The peak-height rule with a smaller value of  $S(q_c)$ manifests that microcrystallization emerges much more readily in the 2D models than in their 3D counterparts. This seems counter-intuitive since one might notice that the freezing transition of the height of the peak value in 2D hard-sphere systems is almost two times higher~\cite{broughton:1982,hoffmann:2001,wang:2010,wang:2011}. To vindicate our evidence, we extend the interpretation of clusterization of the 3D-SALR system in Ref.~\cite{bolli:2016} to our Q2D-SALR scenario. The freezing rule of hard-sphere (disk) systems is based on dispersions experiencing phase transition on account of excluded volume (\textit{packing}) effect. In contrast, the clustering of SALR systems is dictated by the competitive nature between attraction and repulsion. We observe that an increase of attraction facilitates not only the IRO but also the first neighboring order. Intriguingly, by progressively increasing densities, the packing effect becomes appreciable and preempts the modulated clusterization stemning from SALR interactions . This is revealed by the pronounced first neighbor main peak in $S(q)$ for $\phi_{2D}=0.38$ and $0.5$ whilst the prepeak value due to competing interaction is somewhat suppressed.\\
\indent We now scrutinize the prepeak-width rule by fitting Eq.~\eqref{eq:sqfit}. The derived thermal correlation marks the equilibrium clusters phase, which is in the range of $\xi_T/\sigma\gtrapprox2$. This indicates that the IRO $\xi_T/\sigma$ characterizing the cluster formation exceeds the competing characteristic lengthscale, namely the (normalized) Debye screening length $\lambda/\sigma$, which is fixed at $1.794$ in our study. Our results are similar to what Bollinger and Truskett discussed for 3D-SALR systems in Ref.~\cite{bolli:2016}, albeit our current simulations have not swept a broad range of $\lambda$ values. \\
\indent Therefore, akin to the 3D criteria suggested by both Godfrin and Bollinger, we propose the following \textit{hybrid heuristic} criterion for positing the Q2D-SALR equilibrium clusters phase: (a)~the prepeak height surpasses $S(q_c)\approx1.4$ and (b)~the correlation length $\xi_T/\sigma$ corresponding to the width of the prepeak is not less than $2.0$.
\begin{figure}[h]
	\centering
	\begin{picture}(100,240)
	\put(-70,-50){\includegraphics[height=0.6\textwidth,trim={0 0 0 0},clip]{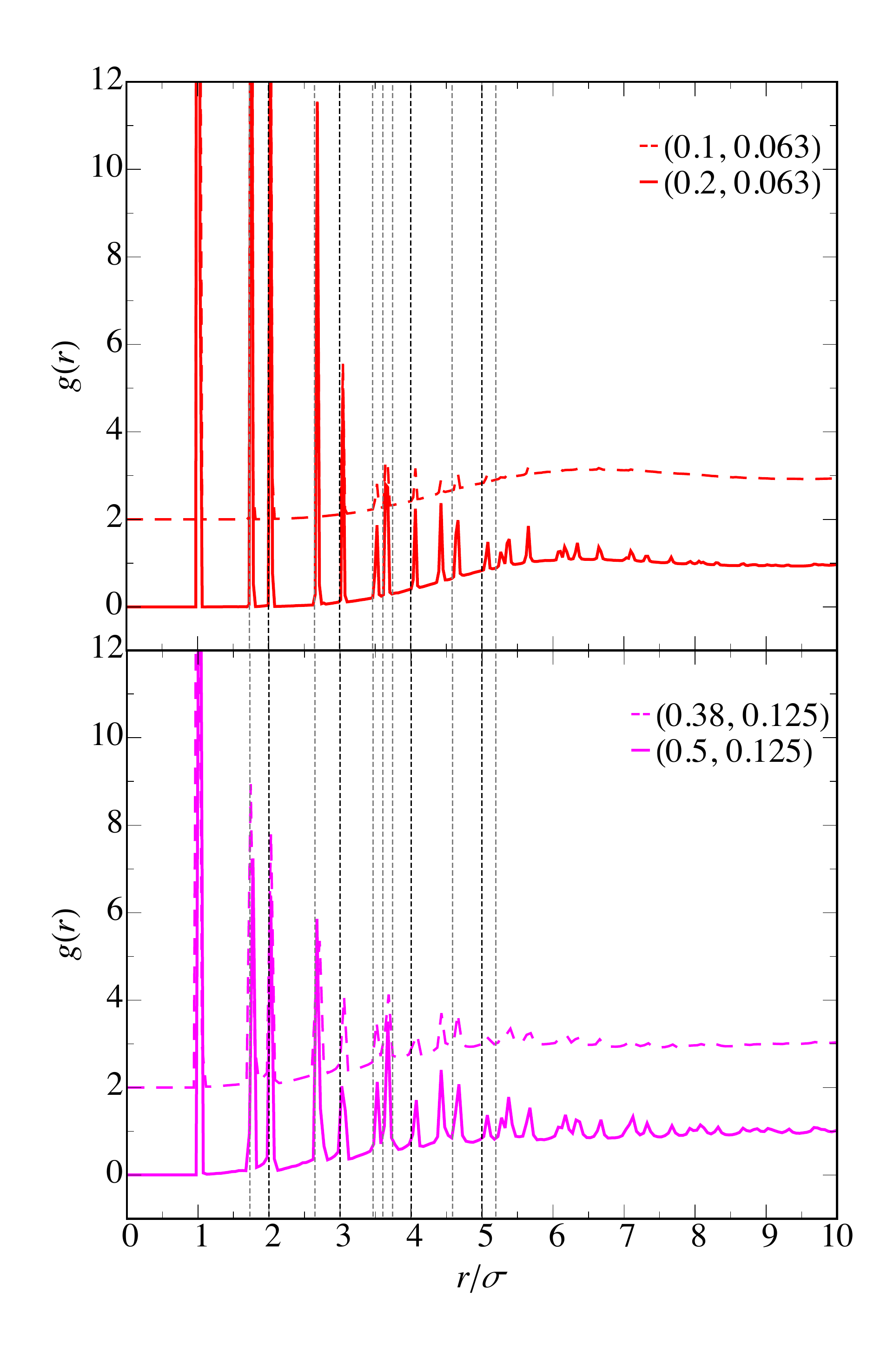}}
	\put(-41,227){(a)}
	\put(-41,100){(b)}
\end{picture}
	\vspace{10mm}
	\caption{\label{fig:g_r} Radial distribution function $g(r)$ for selected systems of SALR particles with different $(\phi_{2D}, T^*)$ values as indicated at (a)~equilibrium clusters and (b)~cluster percolation phases. The positions of all dashed curves are shifted upwards by two units. Vertical dashed lines mark the locations of the peaks for particles packed on a $2D$ hexagonal crystal. (See main text for details.)}
\end{figure}
\begin{figure*}[t]
	\centering
	\begin{picture}(200,300)
		\put(-160,-5){\includegraphics[height=0.6\textwidth,trim={350 200 200 150},clip]{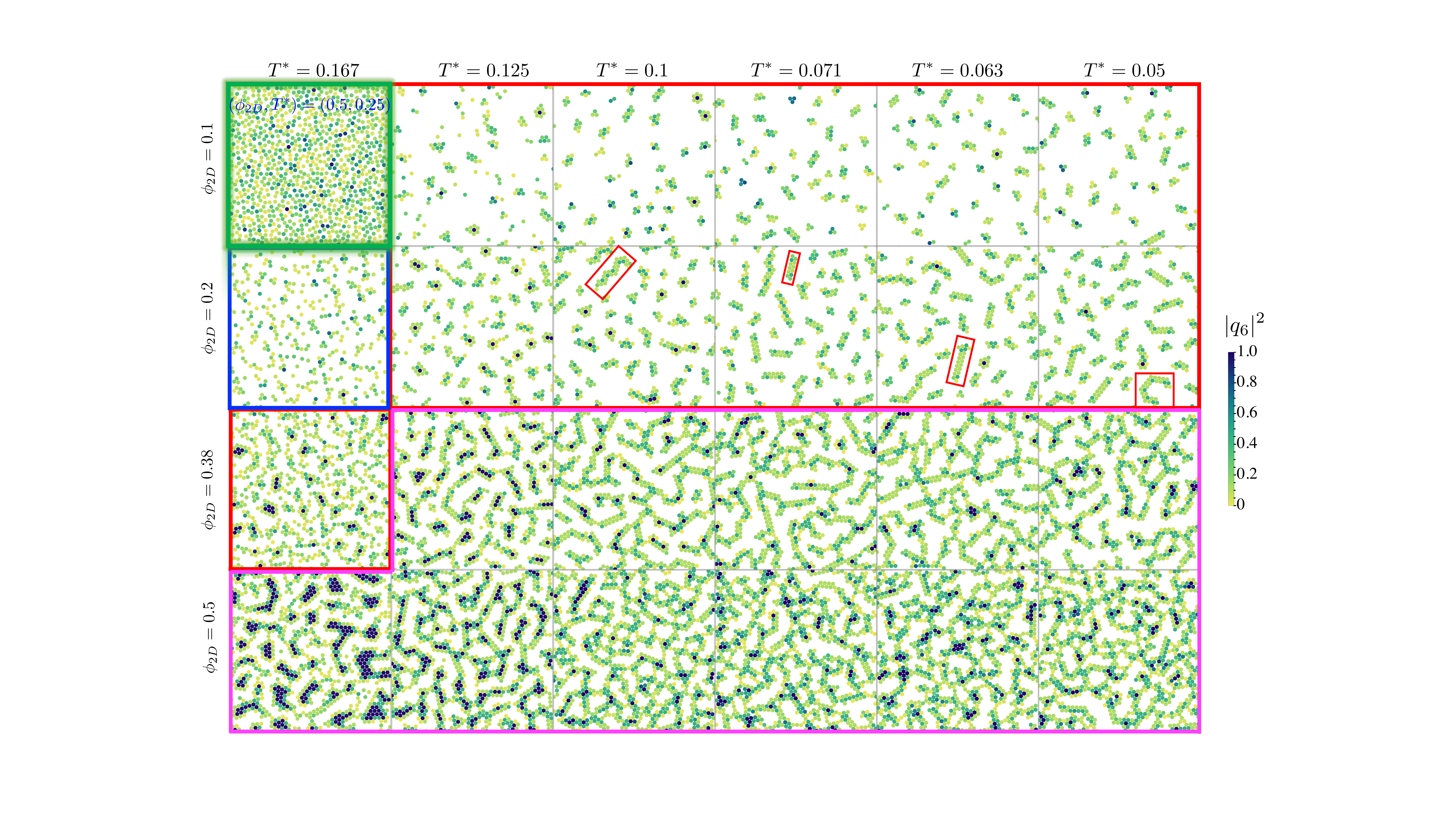}}
	\end{picture}
	\caption{\label{fig:q6sq}Typical configurations of Q2D-SALR systems at different values of density $\phi_{2D}$ and effective temperature $T^*$, except the one at the left top corner corresponding to $(\phi_{2D}, T^*)=(0.5, 0.25)$ labeled in blue, and encompassed by the green line. The local hexagonal order parameter $\left|q_6\right|^2\in [0,1]$ is color-coded. The configurations encompassed by colored frames correspond to different phases classified in Fig.~\ref{fig:phases}. To wit: dispersed fluid (blue), equilibrium clusters (red), random percolation (green), and cluster percolation (magenta). Four double-strand hexagonal clusters (see main text for details) are highlighted by small red rectangles.}
\end{figure*}
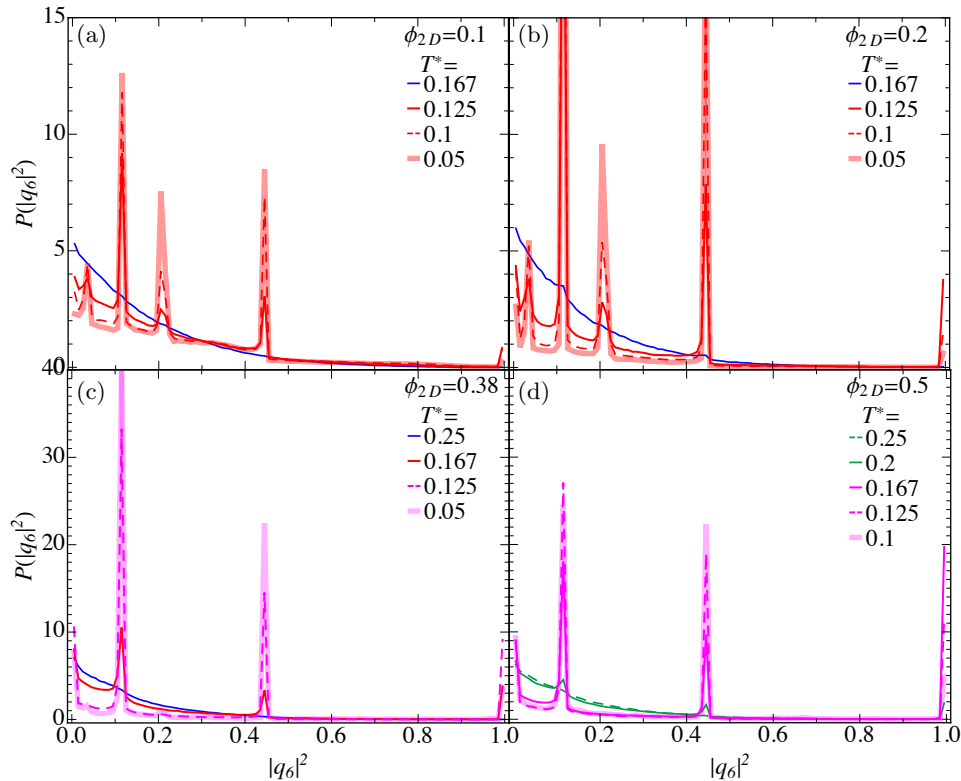
\begin{figure*}[ht]
	\centering
	\begin{picture}(200,280)
		\put(-100,0){\includegraphics[height=0.6\textwidth,trim={0 0 100 100},clip]{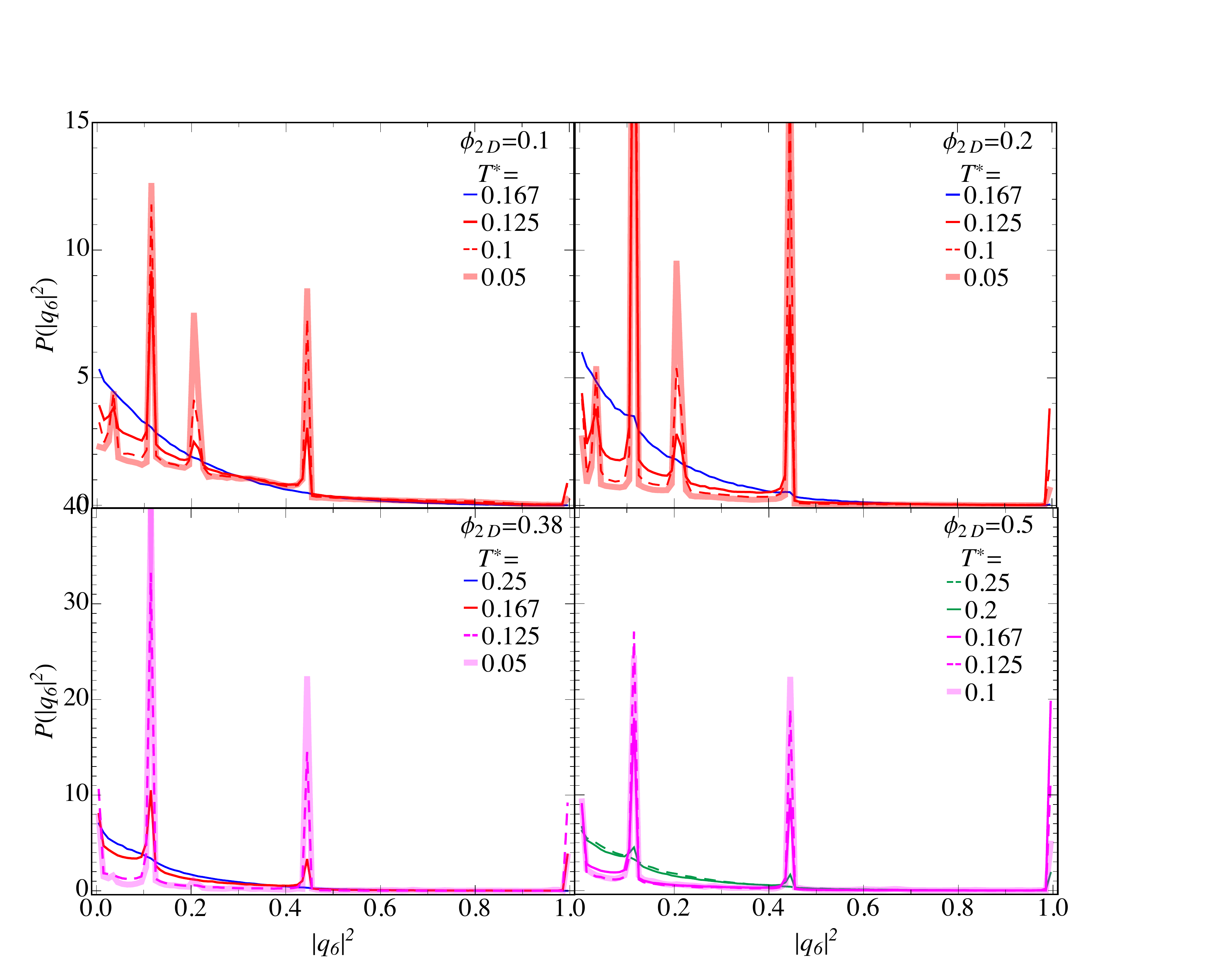}}
		\put(-65,280){(a)}
		\put(102,280){(b)}
		\put(-65,145){(c)}
		\put(102,145){(d)}
	\end{picture}
	\caption{\label{fig:q6}~The probability density function of the hexatic order parameter $P(\left|q_6\right|^2)$ for $\phi_{2D}$ values of (a)~$0.1$, (b)~$0.2$, (c)~$0.38$, and (d)~$0.5$ and for different $T^*$ values as indicated. The colors of the curves are consistent with four characteristic phases as reported in Fig.~\ref{fig:std}. For better visual inspection, some curves are made dashed or transparent.}
\end{figure*}
\begin{figure}[t]
	\centering
	\begin{picture}(100,180)
		\put(-85,-10){\includegraphics[width=0.45\textwidth,trim={0 0 0 0},clip]{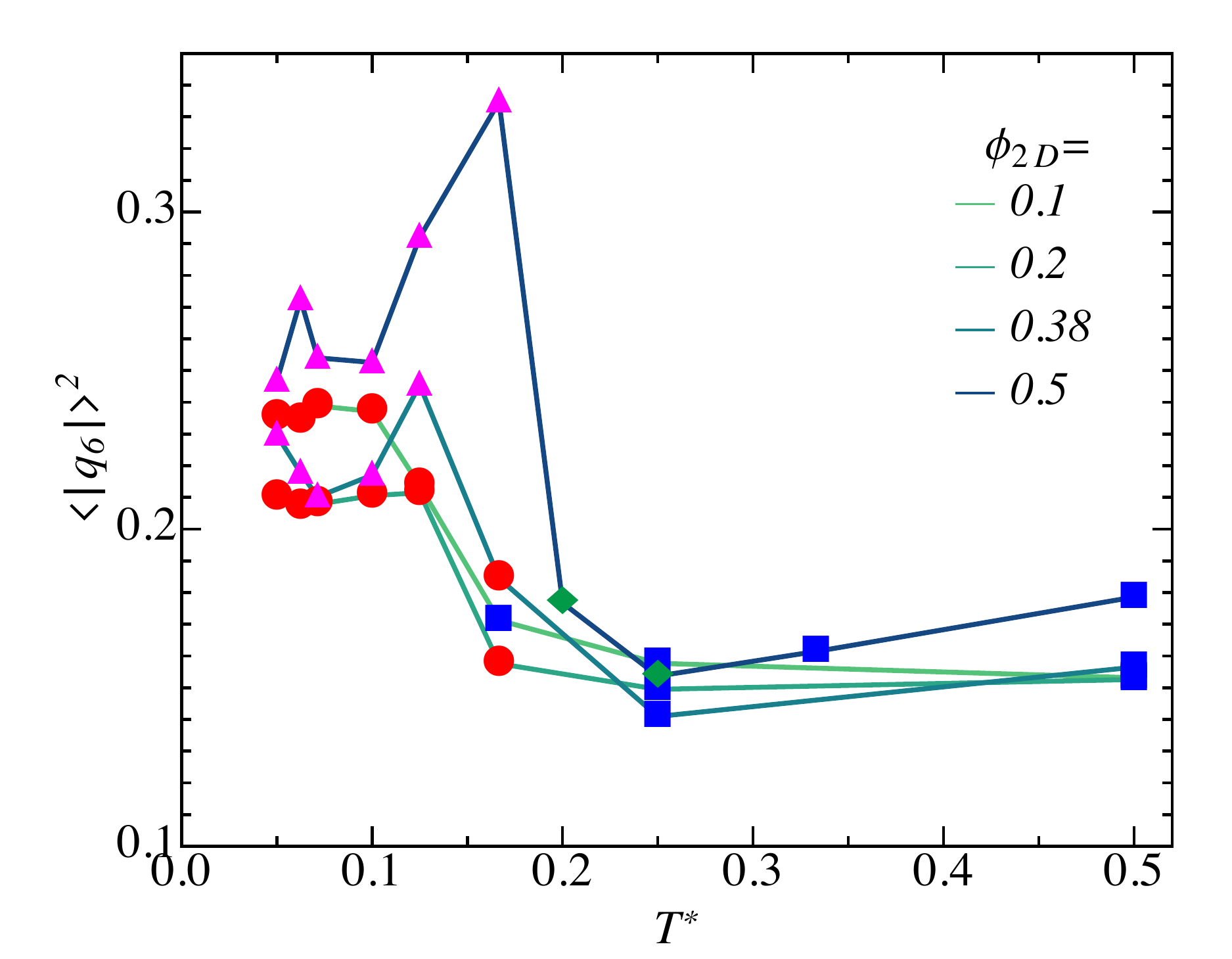}}
	\end{picture}
	\caption{\label{fig:q6_sq}The hexagonal order parameter $|q_6|^2$ as a function of the effective temperature $T^*$.}
\end{figure}
\subsection{Cluster properties}
	\subsubsection{Hexagonal clustering and double-strand hexagonal clusters}
\label{sec:hexel}
Q2D SALR particles preferentially form clusters of finite size hexagonally organized, which are energetically favorable. Indeed, from a simple visual inspection of the equilibrium clusters and cluster percolated phases shown in Fig.~\ref{fig:snapshots}, one can infer an apparent hexagonal order. This can quantitatively be appreciated by scrutinizing the radial distribution function $g(r)$ (related to the static structure factor $S(q)$ by inverse Fourier transformation). Fig.~\ref{fig:g_r} displays the $g(r)$ of Q2D-SALR systems featuring clusterization, i.e. at equilibrium clusters and cluster percolation phases. Two sets of  $(\phi_{2D}, T^*)$ values are considered for each phase. For $(\phi_{2D}, T^*)=(0.1, 0.063)$ the system already develops long-ranged inter particle correlations, as suggested by the  multiple solid-like peaks of the g(r). Sizable peaks persist at distances up to at least five times the particle diameter, thus indicating an ordered arrangement. This feature is more appreciated at $\phi_{2D}=0.2$. More interestingly, as in 3D, in the wake of liquid-like disorder after $r>2\sigma$, the occurrence of  long-range shallow peak at $r\approx6\sigma$ indicates an appreciable cluster-cluster spatial correlation. As shown in Fig.~\ref{fig:g_r} (b), cluster percolated SALR particles develop alike density dependent solid-like mutual particle correlation, whereas the liquid-like shallow peak is flattened. Moreover, a hexagonal clustering feature is encoded in the positions of these peaks, whether these clusters are liquid-like correlated (i.e., equilibrium clusters) or percolated.
The dashed vertical lines in Fig.~\ref{fig:g_r}	are the positions (in unit of $r/\sigma$) of the $g(r)$ peaks of a hexagonal crystal. The latter match very well the peaks of the Q2D-SALR systems (note that agreement persists up to  $r=10\sigma$ for $(\phi_{2D}, T^*)=(0.5, 0.125)$). Explicitly, they are at $r/\sigma=1, \sqrt{3}, 2, 3, 2\sqrt{3}, \sqrt{13}, \sqrt{14}, 4, \sqrt{21}, 5, 3\sqrt{3}...$. \\
\indent To further rationalize how hexagonal clusterization of Q2D-SALR particles depends on effective temperature $T^*$ and density, we calculate the local hexagonal order parameter $\left|q_6\right|^2$ (cf. Eq.~\eqref{eq:hex}) of particles at equilibrated state. Typical configurations for simulations with different parameter pairs $(\phi_{2D}, T^*)$, corresponding to clustered (equilibrium clusters or cluster percolation phases) or close to clustered phases, are sketched in Fig.~\ref{fig:q6sq}. Here the local hexagonal order parameter of each particle is color-coded. The $\left|q_6\right|^2$ map is trivial in the disperse fluid phase ($\phi_{2D}=0.2$ and $T^*=0.167$, enclosed by the thick blue frame) with most of the particles showing $\left|q_6\right|^2\approx0$. So is for random percolation phase (in the glowed green frame at the left top corner, and the blue text indicates the corresponding $\phi_{2D}$ and $T^*$ values) owing to the lack of orientational orders. In contrast, apparent hexagonal order is developed in the equilibrium clusters phase, illustrated in the snapshots marked by red edges for three density cases and a broad range of effective temperature. Here the  $\left|q_6\right|^2$ map shows a noticeable reduction of particles with zero-value order parameter. At dilute equilibrium clusters phase, e.g. for the pair $(0.1, 0.125)$, only few small hexagonally packed clusters form, as reflected by the infrequent number of particles with $\left|q_6\right|^2\approx1$ in the map. Indeed, a substantial portion of unclustered particles exhibit $\left|q_6\right|^2\approx0$. On the other hand, for those clustered particles that have less than $6$ neighbors, intermediate values of $\left|q_6\right|^2$ are measured.\\
\indent Turning to the higher attraction cases ($T^*\leq0.1$), a non trivial clustering process occurs. Indeed, by decreasing the temperature from $0.125$ down to $0.1$, the clusters shape progressively changes from mostly symmetric disk-like to elongated, whilst the number or the size of the clusters do not change significantly. Only from $T^*=0.1$ and downward (cf. $\phi_{2D}=0.1$ and $0.2$) a higher number of elongated clusters emerge. We want to emphasize that this type of cluster is mostly double-strand, namely composed of two strings with local hexagonal arrangement (several of them are highlighted by the red rectangles in Fig.~\ref{fig:q6sq}). In the following discussion, we term these new clusters as ``double-strand hexagonal clusters”.\\
\indent Furthermore, the picture of double-strand hexagonal arrangement of particles is more evident in the cluster percolation phase (marked by the magenta frame). As attraction is strengthened, percolated network undergoes further rearrangement resulting in pieced fragments. This \textit{fragmentation} of the percolated network is more evident in the high density ($\phi_{2D}=0.5$) condition: At $T^*=0.167$, the network is primarily made of particles involved in hexagonal clusters (corresponding to $\left|q_6\right|^2\approx1$). Upon a progressive decrease of $T^*$, the section of the network shrinks, being most of the particles bonded with $4$ neighbors, while the network extension grows.\\
\indent Analogously, the elongated cluster formation was found in the 3D case under two conditions. Toledano et al. discovered such clusters at high densities close to and above percolation with a long-ranged screening length $\lambda=2\sigma$~\cite{toledano:2009}. In the other case, a much shorter screening length (a fraction of $\sigma$) leads to quasi-one-dimensional clustering and the formation of peculiar Bernal spirals, observed both in simulation~\cite{zacca:2005} and experimental~\cite{camp:2005prl} works. In the Q2D case, our double-strand hexagonal clustering is in line with the findings of Ref.~\cite{toledano:2009}. This clustering is explained in terms of inter-cluster interactions favoring anisotropic structures, instead of isotropic ones to lower the total potential energy. On top of that, we should emphasize the effect of strengthened attraction giving rise to the anisotropic clusters even at low densities (far away from the percolation limit). We conjecture that the outer cluster particles preferentially rearrange into elongated shapes to counterbalance the net repulsion due to the intra-clusters particles, thus minimizing the total free energy. Especially at lower $T^*$ conditions, it is energetically more favorable for new particles joining the cluster along the elongated direction since positions along the short axis of the cluster have the higher kinetic barrier~\cite{zhang:2012,mani:2014}.\\
\indent We also present the probability distribution of $\left|q\right|^2$ in Fig.~\ref{fig:q6}.  At high $T^*$ and low $\phi_{2D}$, where systems are in dispersed fluid phase without hexagonal ordering, $P(\left|q\right|^2)$ decays monotonically (cf. blue lines in Fig.~\ref{fig:q6}(a) and (b)). Contrastingly, at low $T^*$ conditions, the emergence of hexagonal clusters in equilibrium cluster phase for $\phi_{2D}=0.1$, or $0.2$, is supported by multiple peaks in $P(\left|q\right|^2)$. The shallow peak developed at $\left|q\right|^2=1$ agrees with a modest quantity of six-bond-neighbor particles illustrated in Fig.~\ref{fig:q6sq}. In addition, more pronounced peaks at $\left|q\right|^2=1/9$, $1/4$, and $4/9$ are detected in $P(\left|q\right|^2)$, encoding those particles dwelling at the edges of the clusters. Strengthening the attraction levers further those peaks due to the outer cluster particles and the elongated cluster shape. In turn, the $\left|q\right|^2=1$ peak is suppressed. An overshot of the peak height ($>20$) of those outer cluster particles is detected in the cluster percolation phase upon increasing attraction. This finding is consistent with the fragmentation of the double-strand hexagonal structures. One might expect higher $P(\left|q\right|^2)$ peak values for higher density conditions, as it is so for the equilibrium clusters phase (cf. Fig.~\ref{fig:q6}(a) and (b)). Conversely, compared with $\phi_{2D}=0.38$, a reduction of the peak height occurs for $\phi_{2D}=0.5$, likely owing to the more fragmented networks. Interestingly, in the equilibrium cluster (EC) phase there is also a certain amount of specific five-particle and six-particle clusters with hexagonal ordering such that  $|q_6|^2=0.233$ and $|q_6|^2=0.198$, respectively. Since the two order parameter values are close to each other, both contribute to the peak in $P(|q_6^2|)$ visible at $|q_6|^2 \approx 0.2$. In the two percolated phases, the clusters are very large and hence this extra peak does not show up. 
Contrastingly, in the random percolated phase, where particles are still thermally agile and can rearrange their relative orientation readily, the distribution of $P(\left|q\right|^2)$ (Fig~\ref{fig:q6}) is nearly identical to that in the dispersed fluid phase. 

The hexagonal order parameter $\langle\left|q\right|^2\rangle$
is plotted in Fig~\ref{fig:q6_sq} against the effective temperature ($T^*$) at various densities. Here, $\langle\left|q\right|^2\rangle$ reveals no clear density dependence since particles belonging to different clusters are not discriminated while computing $\left|q\right|^2$. When represented as an average quantity, the local hexagonal feature has been largely washed out. Notably, a sharp change of $\left|q\right|^2$ occurs between $T^*=0.1$ and $0.2$. Comparing the two cluster percolated states $(\phi_{2D}=0.5, T^*\leq 0.167)$ and $(\phi_{2D}=0.5, T^*\leq 0.125)$, the later has more particles located at the cluster boundaries. This results in a loss of hexagonal order, as demonstrated in Fig.~\ref{fig:q6sq}. 
\begin{figure*}[t]
	\subfloat{		
		\begin{picture}(100,180)
			\put(-148,0){\includegraphics[width=0.46\textwidth,trim={0 0 0 0},clip]{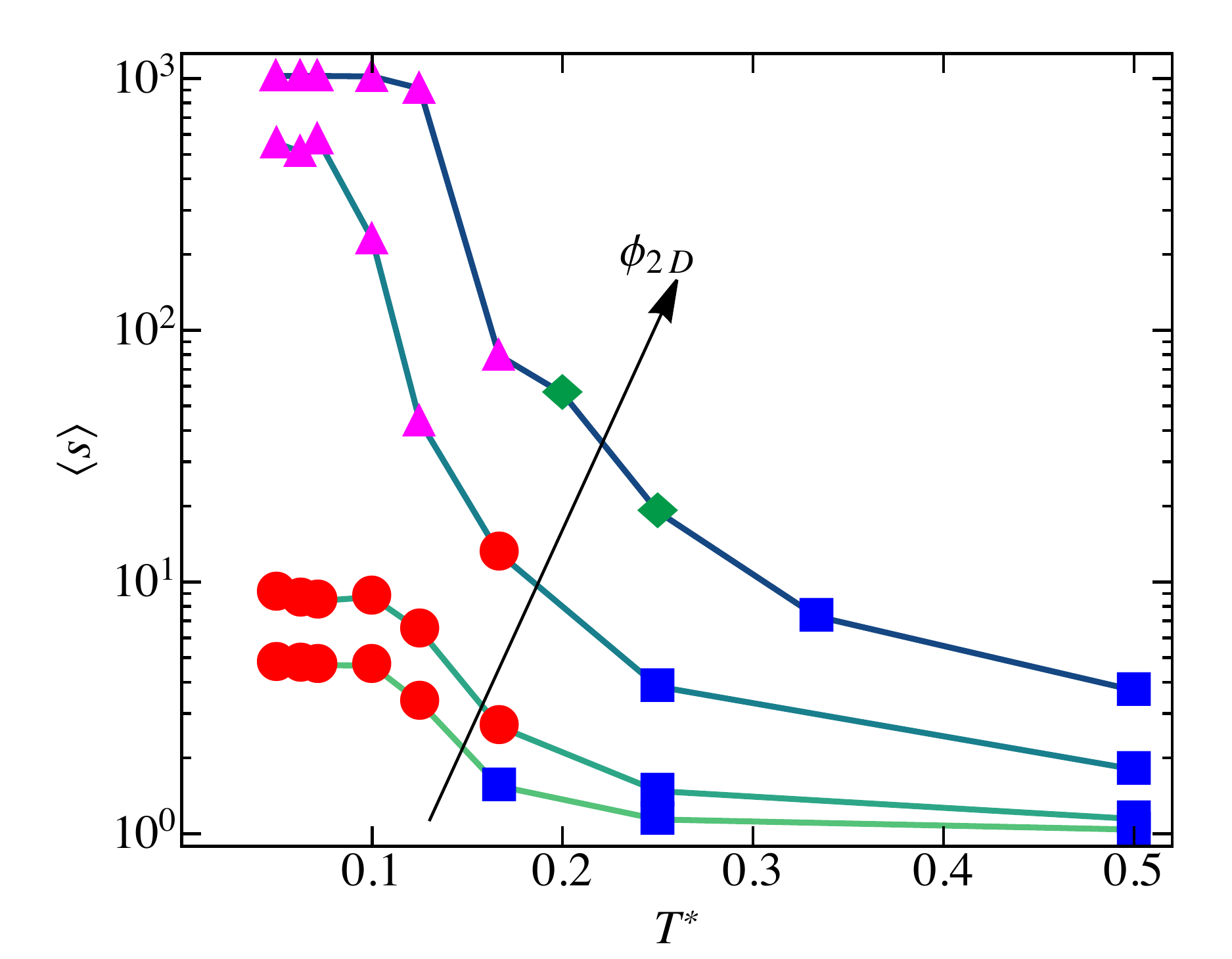}}
			\put(-110,160){(a)}
		\end{picture}
		\label{fig:smax}
	}
	\subfloat{		
		\begin{picture}(100,180)
			\put(-5,0){\includegraphics[width=0.458\textwidth,trim={0 0 0 0},clip]{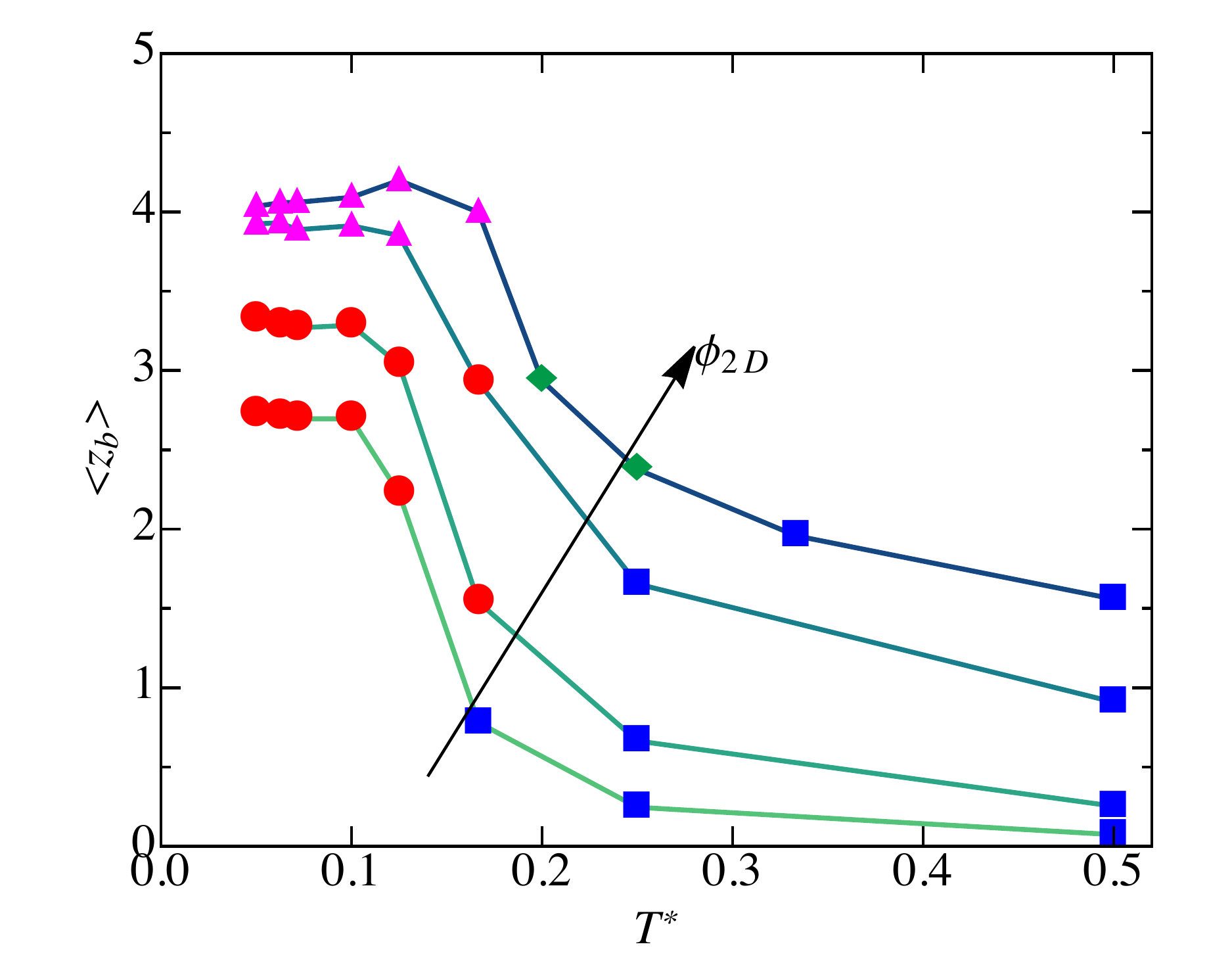}}
			\put(30,160){(b)}
		\end{picture}
		\label{fig:zb_T}
	}\\
	\vspace{2mm}
	\subfloat{		
		\begin{picture}(100,180)
			\put(-150,-0){\includegraphics[width=0.47\textwidth,trim={0 0 0 0},clip]{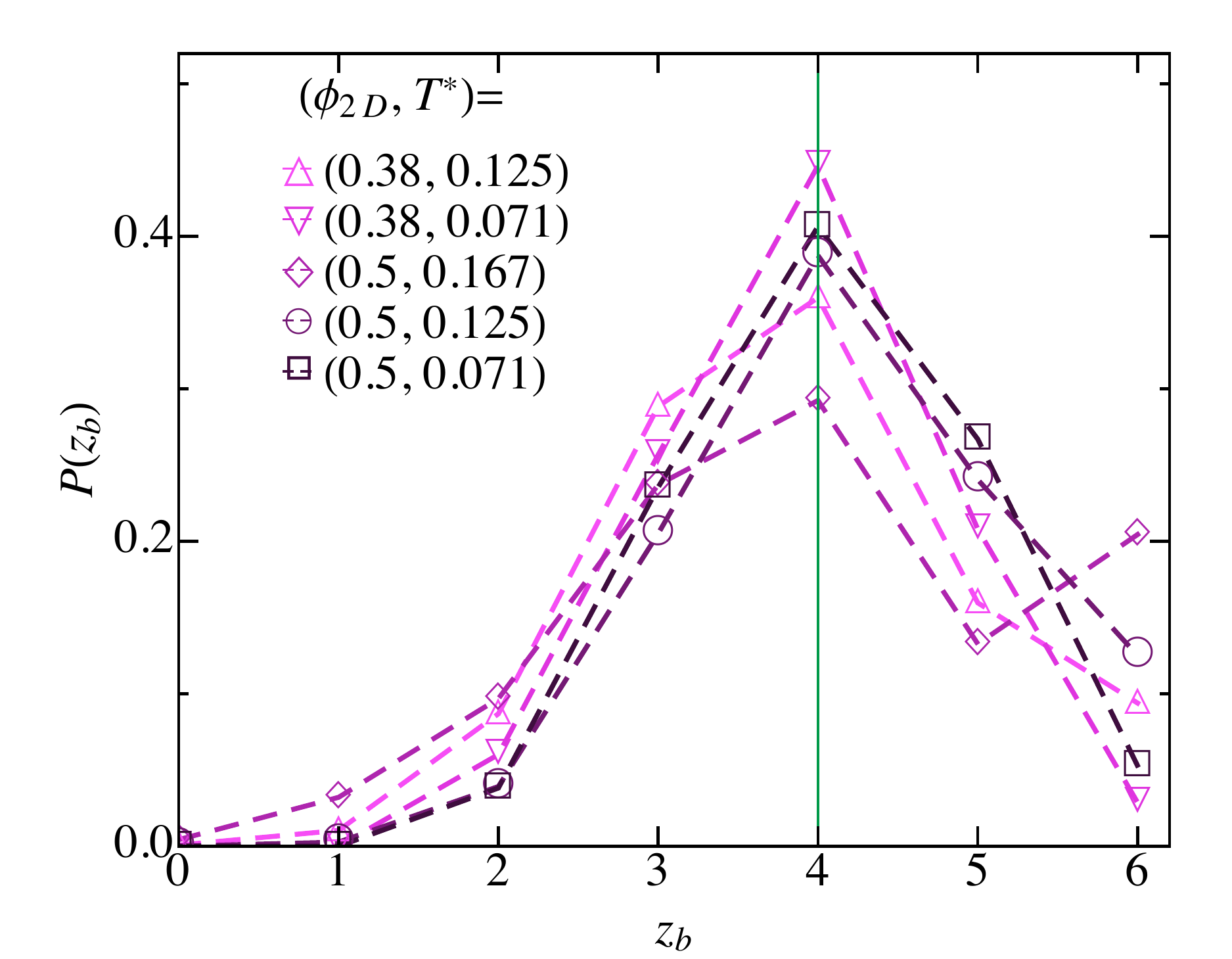}}
			\put(-110,162){(c)}
		\end{picture}
		\label{fig:Pzb}
	}
	\subfloat{		
		\begin{picture}(100,180)
			\put(-3,3){\includegraphics[width=0.46\textwidth,trim={0 0 0 0},clip]{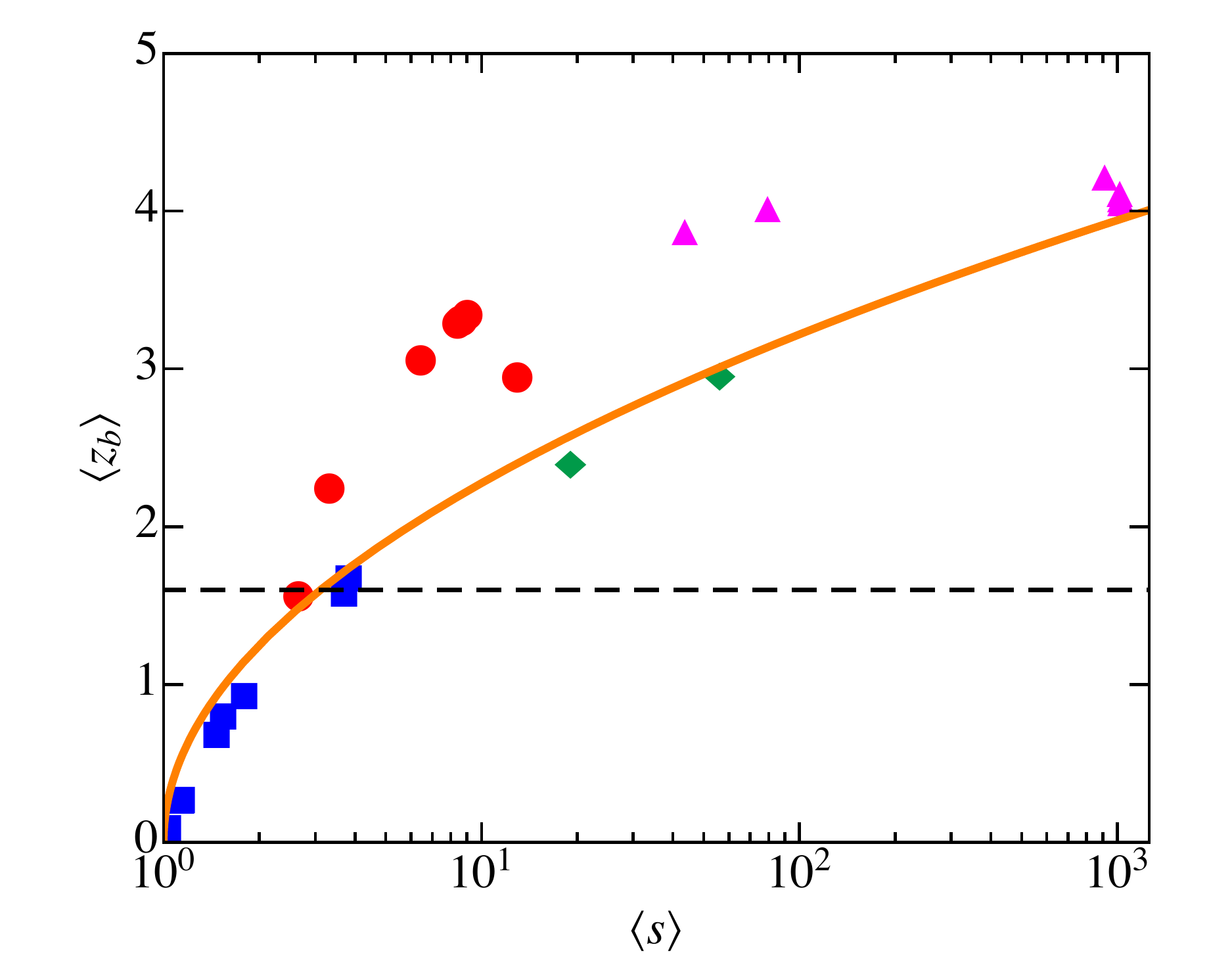}}
			\put(30,160){(d)}
		\end{picture}
		\label{fig:zbs}
	}
	\vspace{3mm}
	\caption{\label{fig:alot}(a)~The mean cluster size $\langle s \rangle$ and (b)~the coordination number (average number of bonds per particle) $\langle z_b\rangle$ as functions of effective temperature $T^*=k_BT/\epsilon$. Dashed arrows are a guide to the eye indicating $\phi_{2D}=0.1$, $0.2$, $0.38$, and $0.5$ in ascending order. (c)~Probability distribution of bond number per particle $P(z_b)$ for cluster percolation (dashed lines and open symbols) phases. (d)~$\langle z_b\rangle$ as a function of average cluster size $\langle s\rangle$ for different $(\phi_{2D}, T^*)$ values. The dashed line marks the critical $\langle z_b\rangle$ value for our simulation data separating disperse fluid and equilibrium clusters phases. The solid orange curve represents the empirical relation in Eq.~\eqref{eq:zbs}. Note that the symbols and their colors correspond to different phases as identified in Fig.~\ref{fig:phases}\protect\subref{fig:std}.}
\end{figure*}
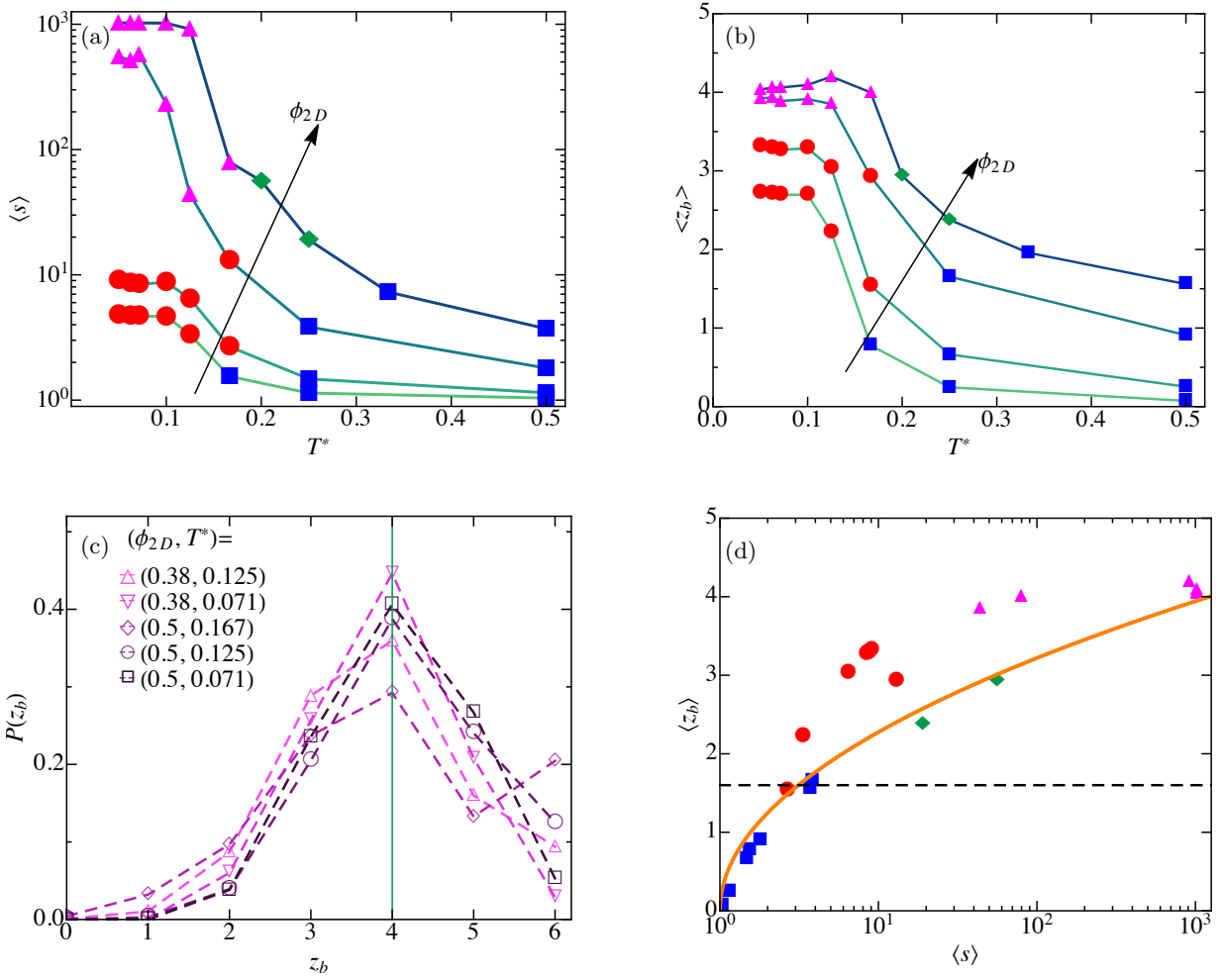
	\subsubsection{Mean cluster size $\langle s\rangle$}
To assess the cluster properties, the $T^*$-dependent evolution of the average cluster size $\langle s\rangle$  at different densities is displayed in Fig.~\ref{fig:alot}\subref{fig:smax}. At first glance, it is observed that $\langle s\rangle$ increases as lowering the effective temperature (strengthening attraction $\epsilon$) and increasing $\phi_{2D}$, as expected. Particularly, $\langle s\rangle\approx N$ for systems at $(\phi_{2D}, T^*)=(0.5, \leq0.125)$, indicative of highly percolated cluster phase. Interestingly, albeit in the dispersed fluid phase, systems at higher density might showcase a larger mean cluster size than that in the equilibrium clusters phase  (e.g., $(\phi_{2D}, T^*)=(0.38, 0.25)$ and $(0.2, 0.125)$), owing to the packing effects induced by the high particle density. However, the clusters in the equilibrium clusters phase are more stable.  Intriguingly, by further ``cooling" the systems, $\langle s\rangle$ shows a \textit{sigmoidal} shape and reaches a plateau whose magnitude depends on the particle density. Taking a closer look at the cluster size distribution, one observes that for $T^*<0.1$ (data not shown), $N(s)$ becomes insensitive to effective temperature and the clusters of size close to the optimal value $\langle s\rangle$ are more likely observed. 

Notice that in the case of 3D studies based on classical nucleation theory,  Zhang et al. ~\cite{zhang:2012} predicted an initial increase of $\langle s\rangle$ when the attraction is strengthened, which is in good agreement with our simulation results for $\epsilon/k_BT\leq10$, whilst for stronger attraction($\epsilon\approx16k_BT$) they reported a shrinkage of $\langle s\rangle$.
\subsubsection{Coordination number $\langle z_b\rangle$}
Insights on the clusters local order can also be obtained by the averaged bond number $\langle z_b\rangle$, also termed as ``coordination” number. The evolution of  $\langle z_b\rangle$ as a function of $T^*$ for different density values is shown in Fig.~\ref{fig:alot}\subref{fig:zb_T}. As explicitly introduced in Sec.~\ref{sec:hexel}, a bond is formed when two particles are closer than $x^*$. In a hexagonal close-packed 2D crystal, $z_b=6$ for each particle. Experimentally, by means of optical microscopy, one can compute $\langle z_b\rangle$~\cite{godfrin:2014}. Loosely speaking, the $\langle z_b\rangle$ data manifests a similar overall trend to $\langle s\rangle$ as shown in Fig.~\ref{fig:alot}\subref{fig:smax}. An increased density leads to a higher $\langle z_b\rangle$ for systems with the same effective temperature, and likewise, $\langle z_b\rangle (T^*)$ curves exhibit a sigmoidal transition from higher to lower temperature and maintain nearly constant $\langle z_b\rangle$ values. We also compute the probability distribution function of the bond number $z_b$, i.e., $P(z_b)$ for different $(\phi_{2D}, T^*)$ values (Fig.~\ref{fig:alot}\subref{fig:Pzb}). For brevity, only $5$ out of $11$ points at the cluster percolation phase (dashed lines and open symbols) are displayed. A local shallow peak at $z_b=6$ arises, witnessing the emergence of hexagonal clusters (see Fig.~\ref{fig:snapshots} for the cluster morphology). In the wake of the development of double-strand hexagonal clusters, a typical scenario of particles surrounded by four neighbors dominates the clustering, resulting in the $P(z_b=4)$ peak. This argument is supported by all the curves in Fig.~\ref{fig:alot}\subref{fig:Pzb} for cluster percolation phase. Specifically, for those lines belong to the same $\phi_{2D}$($0.38$ or $0.5$), a more pronounced peak upon increasing attraction (lowering $T^*$). Notably, among all the state points we simulated, we observe a local maximum of $P(z_b)$ at $z_b=6$ corresponding to the $(\phi_{2D}, T^*)=(0.5, 0.167)$.

Naturally, $\langle z_b\rangle$ is positively correlated with the mean cluster size $\langle s\rangle$~\cite{godfrin:2014,zacca:2005}. Owing to the isotropic central interaction between particles, as clusters grow in size, more particles are included within the cluster, thus showing a higher coordination number than those at the cluster edge or free in solution. This correlation is reported in Fig.~\ref{fig:alot}\subref{fig:zbs} for all the state points studied, with symbols and colors indicating their phases as in Fig.~\ref{fig:alot}\subref{fig:std}. Besides, values in the equilibrium clusters and cluster percolated phases are generally higher than the ones in the dispersed fluid and random percolated phases. When $\langle z_b\rangle$ is above the ``critical value" $1.6$ (the horizontal dashed line in Fig.~\ref{fig:alot}\subref{fig:zbs}), equilibrium clusters and cluster percolation phases emerge, hallmarking a ``critical" mean cluster size $\langle s\rangle\approx4$. Similar to the interpretation provided in Ref.~\cite{godfrin:2014}, a larger $\langle z_b\rangle$ leads to an enthalpic increase of free energy due to SA interactions, which counterbalance the entropic reduction owing to clustering. As a result, the system moves to a lower free energy state with preferred average cluster size $\langle s\rangle$~\cite{godfrin:2014}. Moreover, independent of the spatial dimensionality, an empirical relation given by~\cite{godfrin:2014}
\begin{equation}
	\left< z_b\right> =1.5(\ln\left<s\right>)^{1/2},
	\label{eq:zbs}
\end{equation}
separating the dispersed fluid and random percolated phases from the other two counterparts, still holds in Q2D-SALR systems (solid orange line in Fig.~\ref{fig:alot}\subref{fig:zbs}), indicating that $\langle z_b\rangle$ and $\langle s\rangle$ are similar local order quantities from different perspectives.

Summing up, in this part we have a comprehensive and coherent picture of the cluster properties in Q2D-SALR fluid for a broad range of effective temperatures and densities. The (2D) hexagonal arrangement of the clustering is confirmed. A sufficiently attractive Q2D-SALR system tends to form double-strand hexagonal clusters even at low densities. At higher density, fragmented percolated networks form, whose morphology is still dominated by the double-strand hexagonal pattern.
\section{Summary and Conclusions}
\label{sec:conclusion}
In this work, phase behavior and structural properties of Q2D-SALR systems modeled by Brownian spheres confined in-plane, interacting via the generalized Lennard-Jones–Yukawa (LJY) pair potential for a wide range of effective particle density and effective temperature are systematically investigated. \\
\indent Firstly, we construct a generalized phase diagram by analyzing the cluster size distribution function. This phase diagram bears some resemblance to the one in 3D, wherein we observe the dispersed fluid, equilibrium clusters, random percolation, and cluster percolation phases. We also compare this phase diagram to an attractive SW system with a similar attraction range, using a 2nd-order perturbation theory approach. The binodal curves delineate the boundary between the fluid and clustered phases in the 2D-SALR phase diagram. However, for the parameter range we computed, we did not observe the 2D version of the Noro \& Frenkel-like ELCS, as the normalized second virial coefficient is highly sensitive to the SA range. This deviation from the 3D-SA system may also stem from the second-order perturbation theory used for calculating the binodal in 2D, which is less accurate than that in 3D.\\
\indent Next, we analyze the intermediate-range order peak ($S(q_c)$) that appears in the structure factor for Q2D-SALR systems. We have identified critical values for both the height, analogous to the 3D Hansen-Verlet freezing criterion, and the width, associated with the thermal correlation length $\xi_T/\sigma$, similar to the approach used in 3D. However, qualitatively, the observed critical value $S(q_c)\approx1.4$, which distinguishes the dispersed fluid and equilibrium clusters phases, is only half of that in 3D. This implies that the typical microcrystallization of SALR particles occurs much more readily in the (Q)2D models than in their 3D counterparts. This spatial dimension-dependent criterion is in contrast to the melting transition for hard-sphere (disk) systems. This difference may result from their distinct clustering mechanisms. The former is governed by the competition between attraction and repulsion, while the latter is dictated by packing, specifically the excluded volume effects.\\
\indent The aggregation of (Q)2D-SALR particles is energetically more favorable when they pack in a hexagonal fashion. Consequently, we conducted an analysis of local hexagonal ordering for all investigated states. We observed that as attraction is strengthened (typically up to more than $10k_BT$) at low densities, the equilibrium clusters with finite sizes undergo a transition from spherical-like to elongated shapes. This transition can also be observed at higher densities when the system percolates. Specifically, in Q2D, the anisotropic clusters we found are mainly composed of two linear strings arranged on a hexagonal lattice. This type of 2D cluster bears resemblance to the Bernal spiral observed in 3D, either due to a short screening length or intra- and inter-cluster interactions in the case of a larger screening length.\\
\indent Furthermore, as strengthening the attraction (lowering the effective temperature) or enlarging the density, the mean cluster size grows. However, its dependence on attraction exhibits a sigmoidal increase: It grows gradually for the lower attraction, followed by a sharp increase for intermediate attraction, then reaches a saturated mean cluster size for highly attractive particles. The coordination number follows a similar tendency as the mean cluster size, with a critical value $\langle z_b\rangle\approx1.6$ separating the disperse fluid and equilibrium clusters phases. This critical $z_b$ is analogously in agreement with that in 3D-SALR systems consistent with rigidity percolation threshold ($z_b=2.4$) for 3D covalent glass~\cite{godfrin:2014,HeThorpe:1995}. Moreover, the probability distribution of the bond number $P(z_b)$ suggests that the clustered (i.e., equilibrium clusters and cluster percolation) phases show their peaks at the range $z_b\geq2$. Notably, if the cluster morphology is dominated by double-strand hexagonal clusters or percolation, a peak in $P(z_b)$ persistently occurs at $z_b=4$. Intriguingly, the coupling between the coordination number and the mean cluster size can be agreeably captured by the empirical relation corroborated for 3D-SALR systems signaling a critical mean cluster size $\langle s\rangle\approx4$. \\
\indent To this end, our first comprehensive examination of the structure, phase behavior, and thermodynamics of Q2D-SALR systems may provide insights into the clustering properties of SALR particles confined to interfaces and membrane protein aggregation at the intracellular level.
A natural extension of this study involves investigating the clustering dynamics, including the role of hydrodynamic interactions. This direction is already under preparation in a subsequent paper~\cite{tan05}. Furthermore, it will be intriguing to incorporate the Q2D-SALR system into a fluid-fluid interface and explore the impact of apparent viscosity contrast~\cite{tan03}, given its biological relevance to the interaction of membrane proteins.
\section*{Acknowledgements}
ZT thanks Shibananda Das (UMass Amherst), Kai Qi (2020 X-Lab) and Roland G. Winkler (FZJ) for sharing some cluster analysis scripts and insightful discussions. The authors gratefully acknowledge the computing time granted through JARA-HPC on the supercomputer JURECA at Forschungszentrum J\"ulich~\cite{jureca}.
%

\end{document}